\documentclass[
showkeys,amsmath,amssymb,
nofootinbib,
twocolumn,
floatfix,
superscriptaddress
]{revtex4}
\usepackage{scrextend}
\usepackage{graphicx}
\usepackage{times}
\usepackage{verbatim}
\usepackage{hyperref}
\usepackage[none]{hyphenat}

\begin{document}
%\linenumbers

\title{The Potential to Probe Solar Neutrino Physics with LiCl Water Solution}

\newcommand{\DEP}     {\affiliation{Department~of~Engineering~Physics, Tsinghua~University, Beijing 100084, China}}
\newcommand{\KeyLab}  {\affiliation{Key Laboratory of Particle \& Radiation Imaging (Tsinghua University), Ministry of Education, Beijing 100084, China}}
\newcommand{\Center}  {\affiliation{Center for High Energy Physics, Tsinghua~University, Beijing 100084, China}}
\newcommand{\UCAS}    {\affiliation{School of Physical Sciences, University of Chinese Academy of Sciences, 100049, China}}
\newcommand{\NAN}     {\affiliation{School of Physics, Nanjing University, Nanjing 210093, China}}

\author{Wenhui Shao}\DEP\KeyLab\Center
\author{Weiran Xu\footnote{Now at Laboratory for Nuclear Science, Massachusetts Institute of Technology, MA 02139, USA}}\DEP\KeyLab\Center
\author{Ye Liang}\DEP\KeyLab\Center
\author{Wentai Luo}\UCAS
\author{Tong Xu}\DEP\KeyLab\Center
\author{Ming Qi}\NAN
\author{Jialiang Zhang}\NAN
\author{Benda Xu}\DEP\KeyLab\Center
\author{Zhe Wang\footnote{Corresponding author: wangzhe-hep@mail.tsinghua.edu.cn}}\DEP\KeyLab\Center
\author{Shaomin Chen}\DEP\KeyLab\Center

\date{\today}% It is always \today, today,

\begin{abstract}
Lithium chloride water solution is a good option for solar neutrino detection.
The $\nu_e$ charged-current (CC) interaction cross-section on $\rm{{}^{7}Li}$ is evaluated with new B(GT) experimental measurements.
The total CC interaction cross-section weighted by the solar $^8$B electron neutrino spectrum is $3.759\times10^{-42}~\rm{cm}^2$,
which is about 60 times that of the neutrino-electron elastic scattering process.
The final state effective kinetic energy after the CC interaction on $\rm{{}^{7}Li}$ directly reflects the neutrino energy, which stands in sharp contrast to the plateau structure of recoil electrons of the elastic scattering.
With the high solubility of LiCl of 74.5~g/100~g water at 10$^\circ$C and the high natural abundance of 92.41\%,
the molarity of $\rm{{}^{7}Li}$ in water can reach 11~mol/L for safe operation at room temperature.
The CC event rate of $\nu_e$ on $\rm{{}^{7}Li}$ in the LiCl water solution is comparable to that of neutrino-electron elastic scattering.
In addition, the $\nu_e$ CC interaction with the contained $\rm{{}^{37}Cl}$ also contributes a few percent of the total CC event rate.
The contained $\rm{{}^{35}Cl}$ and $\rm{{}^{6}Li}$ also make a delay-coincidence detection for electron antineutrinos possible.
The recrystallization method is found to be applicable for LiCl sample purification.
The measured attenuation length of $11\pm1$~m at 430~nm shows that the LiCl solution is practicable for a 10-m diameter detector for solar neutrino detection.
Clear advantages are found in studying the upturn effect of solar neutrino oscillation, light sterile neutrinos, and Earth matter effect. The sensitivities in discovering solar neutrino upturn and light sterile neutrinos are shown.
\end{abstract}


\maketitle

\section{Introduction}
The propagation of solar neutrinos has several special features.
At high energy, the $\nu_e$ survival probability is low and dominated by the matter effect, i.e., the MSW effect~\cite{Wolfenstein, Mikheyev},
while at low energy, the probability is high, and the flavor change occurs as in vacuum.
Between the high- and low-energy regions, there is a smooth ``upturn'' of the survival probability.
When arriving at the Earth, they are decoherent mass eigenstates.
The survival probability of $\nu_e$ arriving at a terrestrial experiment is further modulated according to their path in the Earth,
and in the first order, it shows a day-night asymmetry.
The upturn and Earth matter effects are poorly constrained by the SNO~\cite{SNO, SNOThesis} and Super-Kamiokande experiments~\cite{SK, SKTalk}.
In addition to confirming these theoretical predictions, future precise solar neutrino experiments~\cite{HK, LENA, THEIA, Jinping} are expected to probe new physics.
The weakly mixed light sterile neutrino model can influence the upturn curve of the $\nu_e$ survival probability and make it ``dip and wiggle'' in the expected upturn region~\cite{Sterile1, Sterile2}.
Nonstandard interaction (NSI)~\cite{NSI} and light dark matter~\cite{Lopes} are also interesting solar neutrino physics topics to investigate.

The charged-current (CC) interaction of $\nu_e$ on nuclei is most favorable for such types of physics studies because the recoil electron energy is
strongly correlated with the incident neutrino energy.
In radiochemical neutrino experiments with the CC reactions, such as the Homestake experiment~\cite{homestake} with $\rm{{}^{37}Cl}$, GALLEX/GNO~\cite{GALLEX,GNO}, and SAGE~\cite{SAGE} experiments with $\rm{{}^{71}Ga}$,
the energy of the recoil electrons, however, is not measured.
The SNO experiment~\cite{SNO} measured the solar neutrino oscillation with heavy water, and
it is the only experiment by now to give a real-time energy measurement of the CC interactions on nuclei, i.e., deuterium.
However, the CC signals are obscured by the products of neutrino neutral-current (NC) interactions on deuterium because of their similar energy, position, and angular distributions.
Other experimental attempts can also be found for $\rm{{}^{11}B}$~\cite{boron} and $\rm{{}^{115}In}$~\cite{indium}, as well as
a few recent ideas for solar neutrino detection methods with $\rm{{}^{116}Cd}$~\cite{Zuber}, $\rm{{}^{71}Ga}$~\cite{Wang}, $\rm{{}^{131}Xe}$, $\rm{{}^{136}Xe}$~\cite{Haselschwardt}, and others~\cite{Yoshi2}.

The feasibility of using lithium-7 as a target for neutrino detection was recognized in references~\cite{Bahcall:1964, Reines:1965, Kuzmin:1965, Bahcall:1969} in the 1960s.
Experimentally, researchers have proposed to detect the recoil electron signals in an electronic detector with a water solution of lithium chloride~\cite{Peak:1980xb} or to extract the final state $^7$Be signal in radiochemical methods~\cite{Li7Chem, Li7Chem2}.
Recently, the large cross-section is updated and highlighted again~\cite{Haxton}, and
by a solar angle cut, the interaction products of $\nu_e$ with $\rm{{}^{7}Li}$ can be distinguished
from the neutrino elastic scattering on electrons~\cite{ASDC, SNO}.

In this work, we discuss that the detection approach with LiCl water solution is practical and efficient
to explore energy-dependent solar neutrino physics.
In Sec.~\ref{sec:basic}, we present the detection channels of neutrinos in LiCl water solution and, in particular, the CC process on $\rm{{}^{7}Li}$. We present a cross-section estimation of $\nu_e$ CC interaction $\rm{{}^{7}Li}$ with new experimental transition matrix element inputs.
In Sec.~\ref{sec:detector}, we provide our recent investigation of the properties of LiCl water solution and
a compact detector proposal with estimated detector performance.
In section~\ref{sec:advantage}, we clarify the advantage of LiCl water solution in measuring the solar neutrino upturn effect,
the search for light sterile neutrinos, and the study of the Earth matter effect.
The sensitivity of probing the upturn and sterile neutrinos with
the detector proposal is presented in section~\ref{sec:sensitivity}.
The paper concludes in section~\ref{sec:conclude}.

\section{MeV neutrino detection in LiCl water solution}
\label{sec:basic}
In this section, we introduce the detection channels of~MeV neutrinos in LiCl water solution.
The detection of $\nu_e$ is discussed first and then followed by the cross-section calculation of the CC interactions of $\nu_e$ on $\rm{{}^{7}Li}$.
The result is compared with the CC interactions of $\nu_e$ on $\rm{{}^{37}Cl}$ and the elastic scattering of $\nu_x$ on electron,
where $\nu_x$ represents $\nu_e$, $\nu_\mu$, and $\nu_\tau$.
The $\bar\nu_e$ detection is discussed in the end.

\subsection{Detection of neutrinos}
The dominant interactions of neutrinos of 1\text{-}20~MeV in LiCl water solution
are (1) the CC process of $\nu_e$ on $\rm{{}^{7}Li}$ (LiCC),
(2) the CC process of $\nu_e$ on $\rm{{}^{37}Cl}$ (ClCC),
(3) the neutral-current process of $\nu_x$ on $\rm{{}^{7}Li}$ (LiNC),
and (4) the elastic scatter of $\nu_x$ on $e^-$ (Elas).

\begin{figure}[!htbp]
\includegraphics[width=0.4\textwidth]{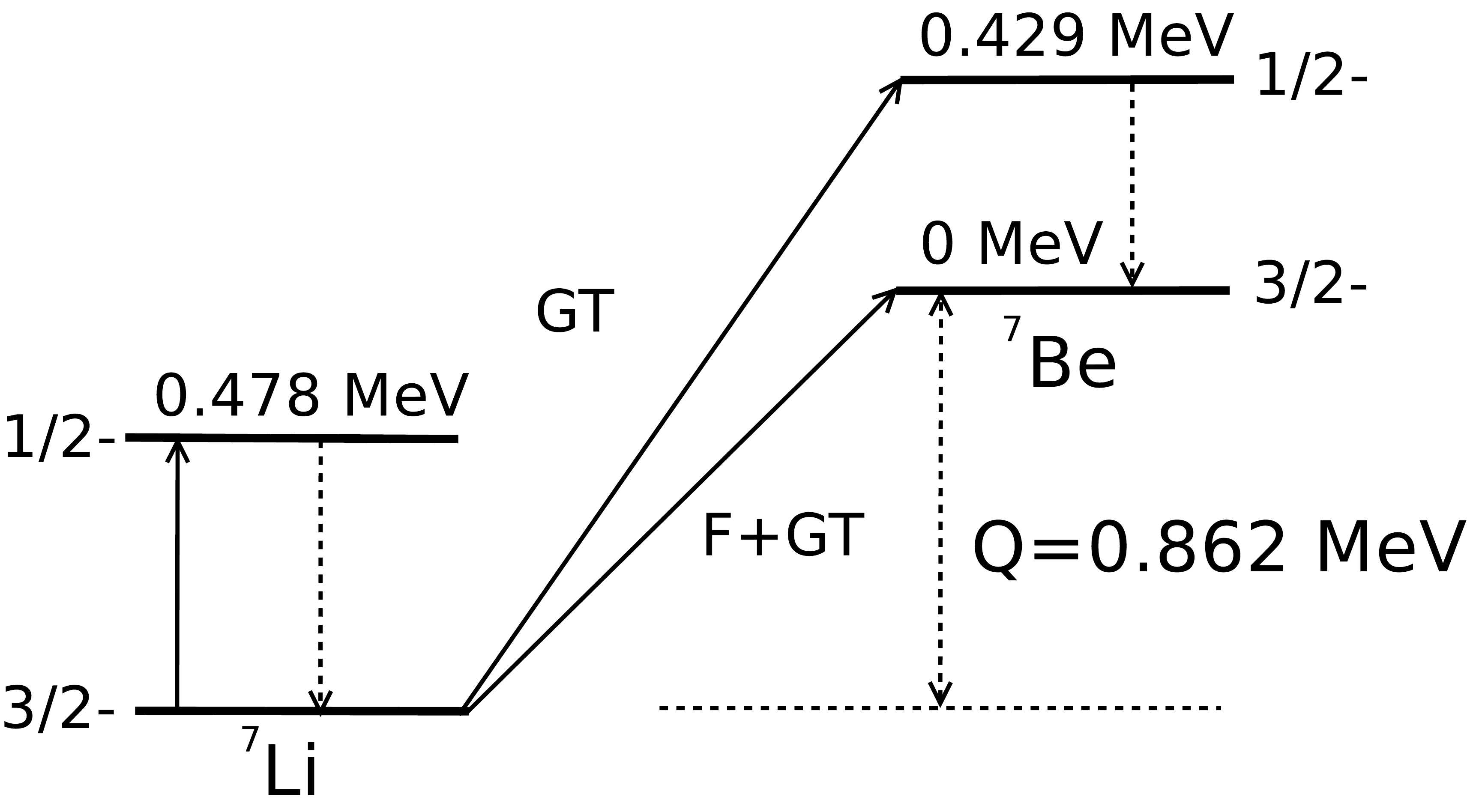}
\caption{\label{fig:levels}
Energy levels of $\rm{{}^{7}Li}$ and $\rm{{}^{7}Be}$ involved in the
neutrino charged current and neutral current interactions on $\rm{{}^{7}Li}$~\cite{Yoshi, Yoshi2}.}
\end{figure}

The LiCC process, as shown in Fig.~\ref{fig:levels}, is
\begin{eqnarray}
\label{eq:LiCC}
\nu_e + {\rm{{}^{7}Li}} \to {\rm{{}^{7}Be}} + e^- (+\gamma).
\end{eqnarray}
The interaction can go through the ground state of $\rm{{}^{7}Be}$ with a threshold of 0.862~MeV, in which both the Fermi and superallowed Gamow-Teller (GT) transitions are possible~\cite{Yoshi, Yoshi2}.
The interaction can also go through the first excited state of $\rm{{}^{7}Be}$ with a threshold of 1.291~MeV, producing an extra 0.429~MeV deexcitation $\gamma$, and the transition is also a superallowed GT transition~\cite{Yoshi, Yoshi2}.

The neutrino energy threshold of the ClCC process is 0.814~MeV, which is very close to the LiCC threshold.
More details of the ClCC process can be found elsewhere~\cite{BahcallBook, Cl37measurement}.

The LiNC process happens as
\begin{eqnarray}
\label{eq:LiNC}
\nu_x + {\rm{{}^{7}Li}} \to \nu_x + {\rm{{}^{7}Li}} + \gamma.
\end{eqnarray}
The energy of the emitted $\gamma$ is 0.478~MeV~\cite{Yoshi, Yoshi2}.
Because it is very low, we skip the discussion involving the NC process in this paper.

The neutrino-electron elastic scattering is
\begin{eqnarray}
\label{eq:Elas}
\nu_x + e^- \to \nu_x + e^-,
\end{eqnarray}
which is one important solar neutrino detection channel~\cite{Borexino, SK-phaseI}.

The final state effective kinetic energy, $T$, of the LiCC process includes the final state $e^-$ and deexcitation $\gamma$ kinetic energy.
It is calculated with the neutrino energy, $E_{\nu}$, as
\begin{eqnarray}
\label{eq:EdetLi}
T = E_{\nu}-{\rm{0.862~MeV}}.
\end{eqnarray}
Similarly, the final state effective kinetic energy of the ClCC process can be calculated as
\begin{eqnarray}
\label{eq:EdetCl}
T = E_{\nu}-{\rm{0.814~MeV}}.
\end{eqnarray}
The $T$ of the CC processes represents the neutrino energy well, and this is critical for energy-dependent physics studies, such as the upturn, sterile neutrinos, and NSI effects.

For the Elas process in Eq.~\ref{eq:Elas}, the kinetic energy, $T$, of the recoil electron shows a plateau structure~\cite{Elastic}, which smooths out many energy-dependent features.

The $T$ of the LiCC, ClCC, and Elas processes is called kinetic energy in the rest of the paper, and it is thought not to cause any confusion.

\begin{figure}[!htbp]
\includegraphics[width=0.5\textwidth]{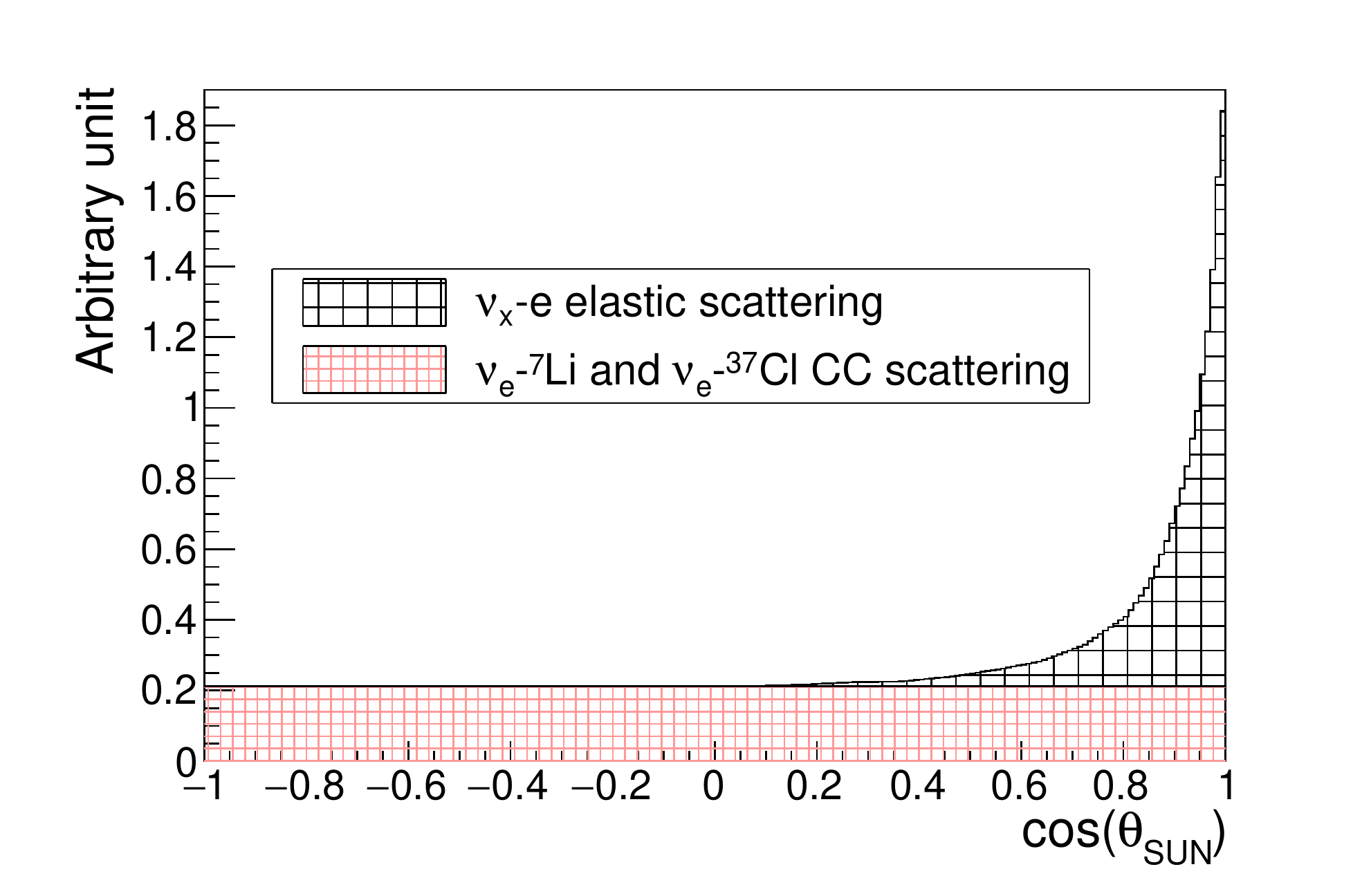}
\caption{\label{fig:angular}
Distribution of the reconstructed solar angles, $\theta_{\rm{Sun}}$, with $T>$5~MeV.
The distribution of $\nu_x$-elastic scattering is extracted from the Super-Kamiokande~\cite{SK-phaseI} and SNO+~\cite{SNOplusTalk} results.
The distribution of $\nu_e$-$\rm{{}^{7}Li}$ and $\nu_e$-$\rm{{}^{37}Cl}$ CC processes is assumed to be uniform.
The ratio of the number of CC events to that of Elas events is set according to the last column of Tab.~\ref{tab:EventRate}.}
\end{figure}

The angular distribution of the recoiling electrons of the LiCC and ClCC processes, as pointed out in~\cite{BahcallBook, Bahcall:1964, Haxton}, is close to a uniform distribution with respect to the incident neutrino direction.
With the contamination of the deexcitation gamma(s) of final states, the reconstructed direction distribution can be even more uniform in a real experiment.
In contrast, the recoil electrons from the Elas process favor the forward direction.
Figure~\ref{fig:angular} shows the distribution of the reconstructed solar angle, $\theta_{\rm{Sun}}$, which is the angle between the reconstructed final particle direction and the solar direction calculated with the position of the Sun.
A uniform distribution is assumed for the CC processes,
and a real distribution for the Elas process is extracted from the Super-Kamiokande~\cite{SK-phaseI} and SNO+ results~\cite{SNOplusTalk}.
Their difference can be used to separate these two types of signals~\cite{SNO, ASDC}.

\subsection{$\rm{{}^{7}Li}$ charged-current cross-section estimation}
\label{sec:cross-section}
\begin{figure}[!htbp]
\includegraphics[width=0.5\textwidth]{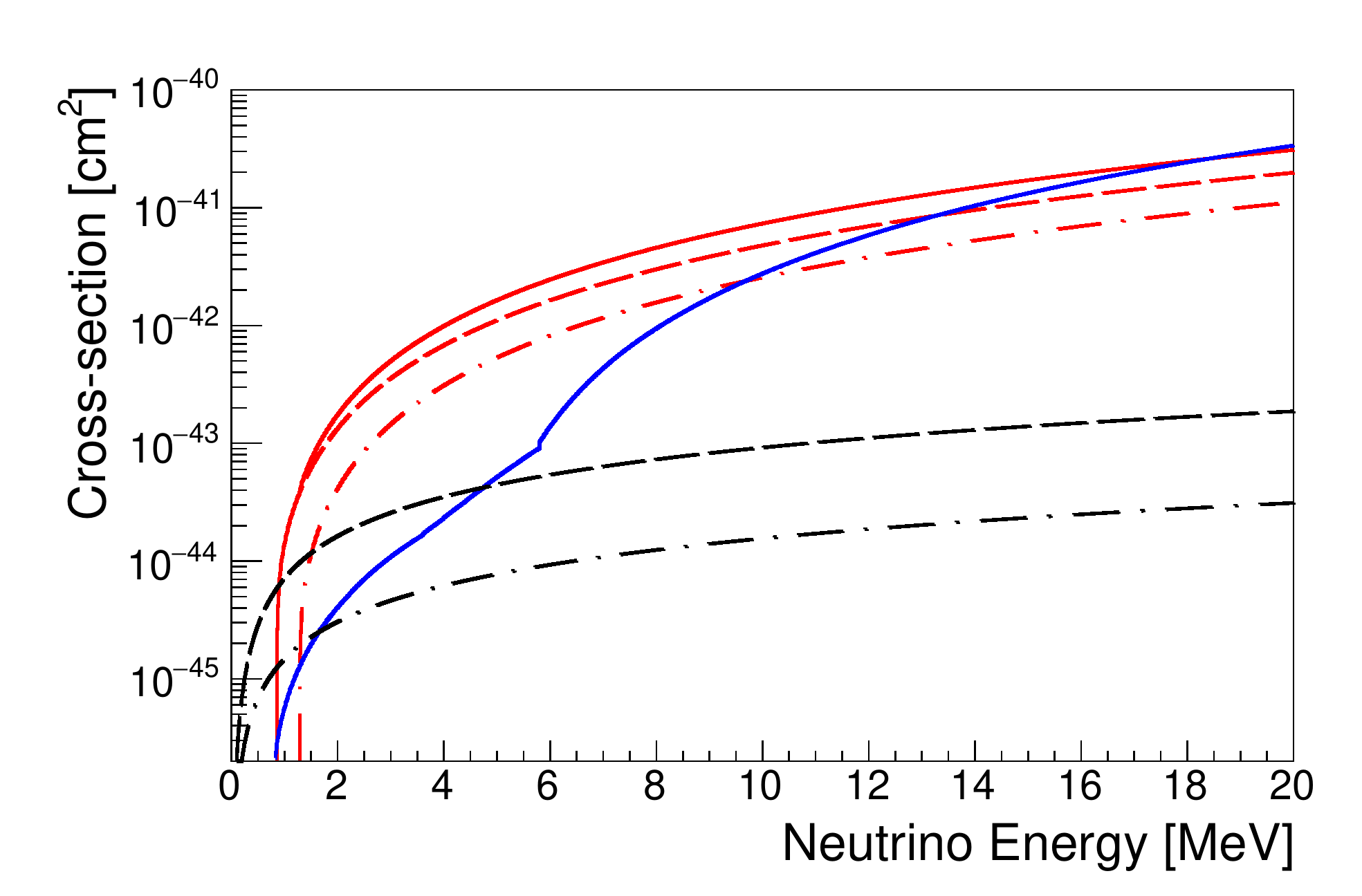}
\caption{\label{fig:CrossSection}
Shown is the cross-section of $\nu_e$ charged-current interaction on $\rm{{}^{7}Li}$ as a function of $\nu_e$ energy, where the total cross-section (red solid), the contribution from the ground state of $\rm{{}^{7}Be}$ (red dashed), and the contribution from the first excited state (red dot-dashed) are shown separately.
The total cross-section of $\nu_e$ charged-current interaction on $\rm{{}^{37}Cl}$ is also shown (blue solid).
The sharp increase at about 5.8~MeV is caused by a Fermi transition to an excited level of $\rm{{}^{37}Ar}$ at 5~MeV.
The elastic scattering cross-sections of $\nu_e$ on $e^-$ (black dashed) and $\nu_{\nu,\tau}$ (black dot-dashed) on $e^-$ are also overlaid for comparison.}
\end{figure}

The LiCC cross-section, $\sigma$, for a specific $E_{\nu}$ and an energy level of $\rm{{}^{7}Be}$ is calculated according to~\cite{BahcallBook, RevModPhys.50.881,Bahcall1966, Barinov:2017ymq}
\begin{align}
\label{eq:Sigma}
\sigma   &= \sigma_0 \frac{\omega_e p_e}{2\pi \alpha Z} F(\omega_e, Z),
\end{align}
where $m_e$ is the electron mass, $\omega_e$ and $p_e$ are the electron energy and momentum in unit of $m_e$, respectively,
$\alpha$ is the fine structure constant, $Z$ is the atomic number of $\rm{{}^{7}Be}$, $F(\omega_e, Z)$ is the Fermi function~\cite{Bahcall1966}, and $\sigma_0$ is
\begin{align}
\label{eq:Sigma0}
\sigma_0 &= \frac{2 \alpha Z m_e^2} {\hbar^4} (G_V^2 \langle1\rangle^2 + G_A^2 \langle\sigma\rangle^2),
\end{align}
where $G_V$ and $G_A$ are the vector and axial coupling constants, respectively, and $\langle1\rangle^2$ and $\langle\sigma\rangle^2$ are the corresponding squares of the Fermi and Gamow-Teller transition matrix elements, represented by B(F) and B(GT) in experimental measurements.
The B(GT) values for the ground and first excited states are measured to be 1.19 and 1.06, respectively~\cite{Yoshi, Yoshi2}.
The energy $\omega_e$ is determined by
\begin{align}
\label{eq:omega}
\omega_e = \frac{E_{\nu} +[M(A,~Z-1) - M(A,~Z)]+ m_e - \bar{E}_{ex}} {m_e},
\end{align}
which depends on $E_{\nu}$, $m_e$, the final atomic mass, $M(A,~Z)$, the initial atomic mass, $M(A,~Z-1)$, and
the average excitation energy of the final atom, $\bar{E}_{ex}$~\cite{RevModPhys.50.881}.
Then, the result is corrected for the screening effect of atomic electrons~\cite{Bahcall1966}.
The cross-section is calculated for the $\rm{{}^{7}Be}$ ground state and the first excited state.
The result as a function of $E_{\nu}$ is shown in Fig.~\ref{fig:CrossSection},
and the total cross-section weighted by an undistorted $^8$B spectrum~\cite{B8spectrum1}
is $3.759\times10^{-42}$~cm$^2$. These quantities for the LiCC cross-section calculation are tabulated in Tab.~\ref{tab:CrossSection}.
\begin{table}[!htbp]
\caption{\label{tab:CrossSection}%
Listed are the energy levels, $E_l$, of $\rm{{}^{7}Be}$ of the
ground level to ground level (gs.-gs.) and ground level to first excited level (gs.-ex.) transitions for the $\nu_e$-$\rm{{}^{7}Li}$ CC process
and the corresponding B(GT) and B(F) values\cite{Yoshi, Yoshi2} and the cross-sections weighted by an undistorted $^8$B neutrino spectrum~\cite{B8spectrum1}.
The total CC cross-section of $\nu_e$-$\rm{{}^{7}Li}$ and $\nu_e$-$\rm{{}^{37}Cl}$ and the $\nu_e$-$e^-$ elastic scattering cross-section are also shown for comparison.}
\begin{ruledtabular}
\begin{tabular}{ccccc}
Channel                 & $E_l$ (MeV) & B(GT)  &  B(F) &  $\sigma(\rm{{}^{8}B})(10^{-42}~\rm{cm}^2)$     \\\hline
$\rm{{}^{7}Li}$ gs.-gs. & 0           & 1.19   &  1    &  2.470                                          \\
$\rm{{}^{7}Li}$ gs.-ex. & 0.429       & 1.06   &       &  1.289                                          \\
$\rm{{}^{7}Li}$ total   &             &        &       &  3.759                                          \\
$\rm{{}^{37}Cl}$ total  &             &        &       &  1.069                                          \\
$e^-$                   &             &        &       &  0.061                                          \\
\end{tabular}
\end{ruledtabular}
\end{table}

Following the same calculation procedure and with the BT strength input from~\cite{Cl37measurement},
the ClCC process cross-section is repeated. The differential cross-section is shown in Fig.~\ref{fig:CrossSection},
and the total cross-section weighted by an undistorted $^8$B spectrum~\cite{B8spectrum1} is $1.069\times10^{-42}$ cm$^2$.
The information is also tabulated in Tab.~\ref{tab:CrossSection}.

Our result for $\rm{{}^{37}Cl}$ is consistent with the result of $1.08\times10^{-42}$ cm$^2$ in~\cite{Cl37measurement} within 1\%,
where the same procedure and input parameters are taken.
However, our new result for $\rm{{}^{7}Li}$ is about 7.5\% higher than those calculated in~\cite{Haxton} and the difference comes from
B(GT) inputs.
The $\nu_e-e$ and $\nu_{\mu,\tau}-e$ elastic cross-sections~\cite{Elastic} are also shown in Fig.~\ref{fig:CrossSection} and Tab.~\ref{tab:CrossSection} for comparison.
For the $^8$B neutrinos, the LiCC cross-section is about 3.5 times that of the ClCC and about 60 times that of the Elas process.

% Does electron capture process also use the following beta- decay ft calculation formula?
%
%The comparable lifetime, $ft$, of the $\rm{{}^{7}Be}$ ground state to $\rm{{}^{7}Li}$
%ground state transition can be calculated with the input of B(GT) from~\cite{Yoshi},
%\begin{align}
%\label{eq:ft}
%ft = \frac{K} {G_V^2 \langle1\rangle^2 + G_A^2 \langle\sigma\rangle^2},
%\end{align}
%with
%\begin{align}
%\label{eq:K}
%K = \frac{2 \ln{2} \pi^3 \hbar^7} {m_e^5 c^4},
%\end{align}
%where $c$ is the light speed. The estimated $ft$ is 2149 s and is 1.9\% higher than the result of ENSDF~\cite{ensdf}, 2109 s.
%The agreement partially validated the new calculation.

A thorough uncertainty analysis is not carried out here, but considering the discrepancy in the neutrino-gallium cross-section validation experiments~\cite{gallium1, gallium2}, we think, in the future, a 2\% uncertainty in the total LiCC cross-section is realistic.

\subsection{Detection of $\bar\nu_e$}
LiCl water solution is also convenient for $\bar\nu_e$ detection.
The water solution contains many hydrogens, i.e., free protons, which are the target of the inverse-beta-decay process of $\bar\nu_e$,
\begin{eqnarray}
\label{eq:IBD}
\bar\nu_e+p \to n + e^+.
\end{eqnarray}
The neutron can be captured on $\rm{{}^{6}Li}$ or $\rm{{}^{35}Cl}$ to form a delayed signal.
The capture cross-sections are 940 barns and 44 barns for $\rm{{}^{6}Li}$ and $\rm{{}^{35}Cl}$, respectively.
The delayed signal is a triton and an $\alpha$ for the capture on $\rm{{}^{6}Li}$~\cite{PROSPECT} and
several gammas with a total energy of 8.6~MeV for the capture on $\rm{{}^{35}Cl}$~\cite{SNO}.
With the natural abundance input of the Li and Cl isotopes, the neutron capture probability on $\rm{{}^{35}Cl}$ is about 30\%,
and the capture probability on $\rm{{}^{6}Li}$ is about 70\%.
The delayed coincidence is excellent in extracting the $\bar\nu_e$ signals and suppressing backgrounds.

\section{Detector with LiCl water solution}
\label{sec:detector}
In this section, we report the $\rm{{}^{7}Li}$, $\rm{{}^{37}Cl}$, and electron molarities in LiCl water solution and the related LiCl purification and the attenuation length of LiCl water solution.
Then, we present a compact detector setup and the corresponding properties, such as the energy resolution, angular resolution, detection threshold, and fiducial volume.

\subsection{$\rm{{}^{7}Li}$, $\rm{{}^{37}Cl}$, and electron molarities in LiCl water solution}
\label{sec:molarity}
A LiCl water solution with high $\rm{{}^{7}Li}$ molarity can be easily achieved at room temperature.
The density of a saturated LiCl water solution at room temperature is measured to be about 1.2~g/cm$^3$~\cite{YeLiang}.
The solubility of LiCl in water is rather high, i.e., 74.5~g/100~g water at 10$^\circ$C, and
Tab.~\ref{tab:solubility} shows the solubility at several temperatures~\cite{chemister}.
The natural abundance of $\rm{{}^{7}Li}$ is 92.41\%~\cite{ensdf}, and the natural abundance of $\rm{{}^{37}Cl}$ is only 24.24\%.
Considering a detector running at room temperature of 20$^\circ$C, to avoid precipitation, we can set the actual
LiCl concentration to be the value at 10$^\circ$C for safe operation.
In such a LiCl water solution, the molarities of $\rm{{}^{7}Li}$, $\rm{{}^{37}Cl}$, and electron are 11, 2.9, and 610~mol/L, respectively, as shown in Tab.~\ref{tab:EventRate}.
\begin{table}[h]
\caption{\label{tab:solubility}%
LiCl solubility in water at several temperatures~\cite{chemister}.}
\begin{ruledtabular}
\begin{tabular}{cc}
Temperature ($^\circ$C)    &  Solubility (g/100~g water) \\\hline
0       &  68.3 \\
10      &  74.5 \\
20      &  83.2 \\
40      &  89.4 \\
60      &  98.8 \\
80      &  112.3 \\
\end{tabular}
\end{ruledtabular}
\end{table}

\begin{table}[!htbp]
\caption{\label{tab:EventRate}%
% Sensitivity/Fit/Li7Upturn/data-table.III
Molarity of $\rm{{}^{7}Li}$, $\rm{{}^{37}Cl}$, and $e^-$ in LiCl solution, in which the LiCl concentration is 74.5~g/100~g water, i.e.,~the saturation solubility at $10^\circ$C.
The event rates for the charged-current interactions of $\nu_e$ on $\rm{{}^{7}Li}$, $\rm{{}^{37}Cl}$,
and the elastic scatterings of $\nu_x$ on $e^-$ are also shown, where
they are calculated with the undistorted $\rm{{}^{8}B}$ $\nu_e$ spectrum~\cite{B8spectrum1}, oscillated spectrum
and oscillated spectrum plus a $T>5$~MeV cut.}
\begin{ruledtabular}
\begin{tabular}{ccccc}
                  & Molarity    & Event rate       & Event rate       & Event rate         \\
                  &             & No osci.         & Osci.            & Osci. \& $>$5~MeV   \\
                  & (mol/L)     & (/100 ton-year)  & (/100 ton-year)  & (/100 ton-year)    \\\hline
$\rm{{}^{7}Li}$   & 11          & 305              & 101              & 87.3               \\
$\rm{{}^{37}Cl}$  & 2.9         & 22.7             & 7.28             & 7.17               \\
All CC            &             & 328              & 108              & 94.4               \\
$e^-$             & 610         & 271              & 124              & 34.5               \\
\end{tabular}
\end{ruledtabular}
\end{table}

\subsection{LiCl purification}
For the application in a neutrino detector, the purification of LiCl is a key question.
For a market sample, usually with a purity of $\ge$99\%, filtration is the first essential procedure to remove the dominant impurity.
Observing that the solubility of LiCl in water varies with temperature, we find that LiCl can be recrystallized and purified by adjusting its saturated solution temperature.
After a round of filtration with a 0.2~$\mu$m membrane and a round of recrystallization,
the potassium, uranium, and thorium concentrations can be suppressed.
More details are given in~\cite{YeLiang}.
%Another common lithium salt lithium carbon carbonate (Li$_2$CO$_3$) can be purified by pumping in sufficient

\subsection{Attenuation length of saturated LiCl water solution}
Our initial test results of the attenuation length and spectrum of a saturated LiCl water solution are satisfactory for a compact neutrino detector.
%The attenuation spectrum of a saturated LiCl water solution at room temperature is measured with a UV spectrometer using a 10-cm long quartz cell.
%The attenuation coefficient is flat from 400 to 700 nm, as shown in Fig.~\ref{fig:Absorption}, and start to rise below 400 nm.
%Below 400 nm, LiCl solution is even better then the popular organic scintillator solution linear alkyl benzene (LAB).
%Due to the long attenuation length, we cannot get a quantified number from the UV spectrometer.
The attenuation length is measured with an 80 cm long tube and an LED.
The emission spectrum of the LED peaks at 430 nm and spans from 375 to 550 nm.
The attenuation length of the sample is measured to be 11$\pm$1~m.
More details of the measurement, the emission spectrum, and the absorption spectrum are given in~\cite{YeLiang}.
We expect that this result can be further improved when higher purification is achieved.
%\begin{figure}[h]
%\includegraphics[width=0.48\textwidth]{figure/LiCl_UV_Absorption.pdf}
%\caption{\label{fig:Absorption}
%Absorption coefficient as a function of wavelength of ultrapure water, saturated LiCl water solution, and linear alkyl benzene (LAB).}
%\end{figure}

\subsection{Neutrino detector proposal and property}
\label{sec:detectorproposal}
With the above information, we see that LiCl water solution can be used for a low background and compact detector.
The detector structure is similar to the SNO (SNO+) experiment~\cite{SNO}.
It has a spherical array of photomultipliers (PMTs), a water buffer, an acrylic vessel,
and LiCl water solution in the center as the neutrino detection medium.
The diameter of the acrylic vessel is 10 meters, which matches the measured attenuation length.
To run the detector safely at 20$^\circ$C, we assume that the LiCl concentration is 74.5~g/100~g water (see the discussion in Sec.~\ref{sec:molarity}).
We assume a photocathode coverage of 50\% and a PMT photon detection efficiency (quantum efficiency times collection efficiency) of 30\%.
Note that LiCl water solution is corrosive to metal. Glass, acrylic, Teflon, or Teflon-lined containers or tools are necessary.
In this work, we focus on the neutrino signals from $^8$B neutrinos.
With the experience from SNO+~\cite{SNOplusTalk} and Super-Kamiokande~\cite{SK} experiments, the energy resolution, angular resolution, fiducial volume, and detection threshold are discussed below.

The Cherenkov light signals from charged particles are detected with the PMTs. The energy resolution for particles follows the Poisson uncertainty of the total number of detected photoelectrons (PEs).
With the higher photocathode coverage and PMT photon detection efficiency,
the total photon detection efficiency is about twice that of the Super-Kamiokande experiment~\cite{SK},
so that a light yield of approximately 20 PE/MeV is expected.

Direction reconstruction is performed with the Cherenkov light.
A resolution of about 35 degrees (68\% C.L.) has been achieved for electrons with energy greater than 5~MeV at the SNO and Super-Kamiokande experiments~\cite{SK, SNO}.
However, for the solar neutrino Elas event study, because the distribution of the reconstructed solar angle, $\theta_{\rm{Sun}}$, is not a simple gauss distribution, the actual distribution from the two experiments is extracted and used in this study, which is shown in Fig.~\ref{fig:angular}.

To reduce the radiative background from PMTs and detector structures, only a central fiducial volume is
available for physics studies.
Because we focus on signals with energy greater than 4 or 5~MeV, the central spherical volume with an 8 m diameter is considered as the fiducial volume for the proposed 10-m diameter acrylic vessel.
The fiducial region is 268~m$^3$ or 320~tons. The total amount of $\rm{{}^{7}Li}$ in the fiducial volume is 2.95$\times10^6$~moles or 20.6~tons.

A 5~MeV detection threshold is realistic for a background free solar neutrino study, as shown in the SNO+ result~\cite{SNOplusTalk}.
Given the better energy resolution in this assumed detector to suppress low energy radioactive or instrumental background, a 4~MeV threshold may also be possible.

We are also interested in a solution with some scintillation component, i.e.,~a water-based liquid scintillator,
in which the scintillation light yield is comparable with Cherenkov light~\cite{ASDC, Biller, Guozy}.
This further enhances the light yield and energy resolution and maintains
a similar performance for the direction reconstruction~\cite{Land,Luowt}.

\section{Advantage in solar neutrino physics study}
\label{sec:advantage}
In this section, we demonstrate that the LiCl-based detection strategy has advantages for energy-dependent neutrino physics studies.
The candidate event rates and spectra of the LiCC, ClCC, and Elas processes are explained first, and then
we look at the solar neutrino upturn issue, the light sterile neutrino search, and the Earth effect.

\subsection{Candidate event rates}
\label{sec:procedure}
\begin{figure}[!htbp]
\includegraphics[width=0.45\textwidth]{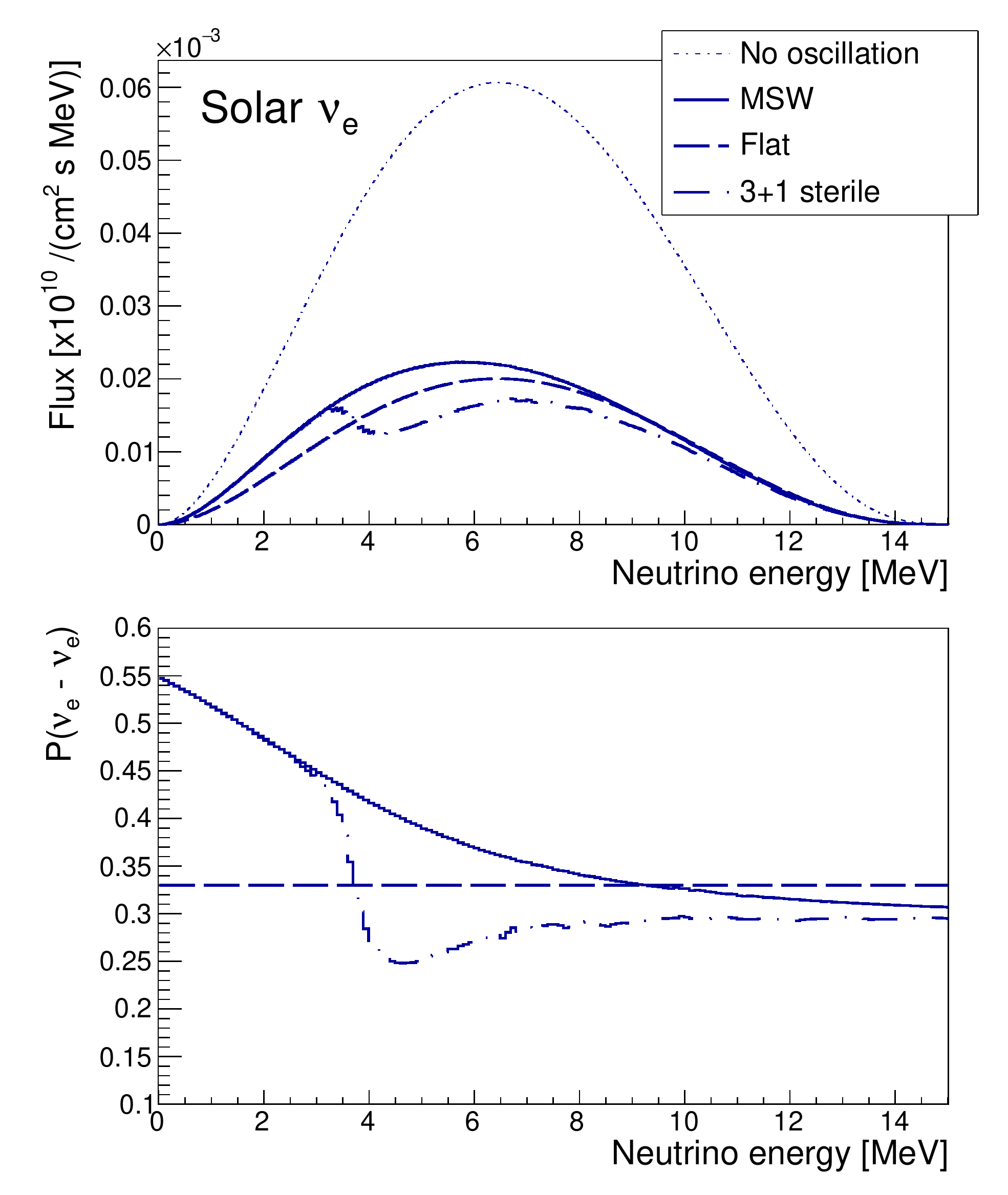} % Here is how to import EPS art
\caption{\label{fig:B8Fig1}
Solar $^8$B electron neutrino energy spectrum with different oscillation configurations (upper panel) and their corresponding survival probability curves (lower panel).
For the sterile neutrino mixing, $\alpha$ = 0.021 and $\Delta m^2_{01}$ = $1.56\times10^{-5}$ eV$^2$.}
\end{figure}

We focus on the $^8$B neutrino studies with a 4- or 5-MeV detection threshold, as explained in section~\ref{sec:detectorproposal}.
In this work, the total $^8$B flux is assumed to be $4.59\times10^{-14}/({\rm{cm^2 s}})$ for the AGS09 low metallicity prediction~\cite{LowMetallicity}, so that
it gives a conservative statistical estimation for sensitivity studies.
We adopt the $^8$B neutrino spectrum prediction, $\Phi_{\nu}(E_{\nu})$, in~\cite{B8spectrum1}, which is shown in Fig.~\ref{fig:B8Fig1}.

The differential energy spectra of all neutrino flavors at a terrestrial detector are evaluated numerically.
The calculation starts with the generation of $^8$B neutrinos according to their spatial probability density function~\cite{NumberDensity}.
Then, their propagation and oscillation probability are estimated.
Their survival or appearance probability of $\nu_e$, $\nu_\mu$ and $\nu_\tau$ at a terrestrial detector is represented with $P_{ee}(E_{\nu})$, $P_{e\mu}(E_{\nu})$ and $P_{e\tau}(E_{\nu})$, respectively.
The calculation methods for the three-active-neutrino propagation in the Sun, the case with one sterile neutrino (3+1), and the neutrino propagation in the Earth are not identical.
Next, the three-active-neutrino propagation in the Sun is introduced first.
More detailed explanations for the (3+1) case and the propagation in the Earth are given in the relevant sections.

For the three-active-neutrino propagation in the Sun, the following calculation is carried out according to the MSW theory.
The $^8$B neutrino generation zone ($r<0.135R_{\odot}$, where $R_{\odot}$ is the radius of the Sun) is divided into 60 shells.
Neutrinos are generated according to the predicted probability density in each shell~\cite{NumberDensity}.
The initial fusion produced $\nu_e$ flux is decomposed into mass eigenstates $\nu_1$, $\nu_2$, and $\nu_3$
according to the local number density of electrons~\cite{NumberDensity}.
With the adiabatic assumption~\cite{Wolfenstein, Mikheyev}, these mass eigenstates are recombined into
flavor eigenstates at the solar surface under vacuum condition.
The neutrino oscillation parameters used are
$\theta_{12}=0.587$~\cite{SK}, $\theta_{13}=0.148$~\cite{DYB}, $\theta_{23}=0.849$~\cite{PDG2020},
$\Delta m^2_{21}=7.49\times10^{-5}~{\rm{eV}^2}$~\cite{SK}, $\Delta m^2_{31}=2.53\times10^{-3}~{\rm{eV}^2}$~\cite{PDG2020}.
The oscillated $^8$B neutrino energy spectrum and $P_{ee}(E_{\nu})$  are also shown in Fig.~\ref{fig:B8Fig1}.

\begin{figure}[!htbp]
\includegraphics[width=0.45\textwidth]{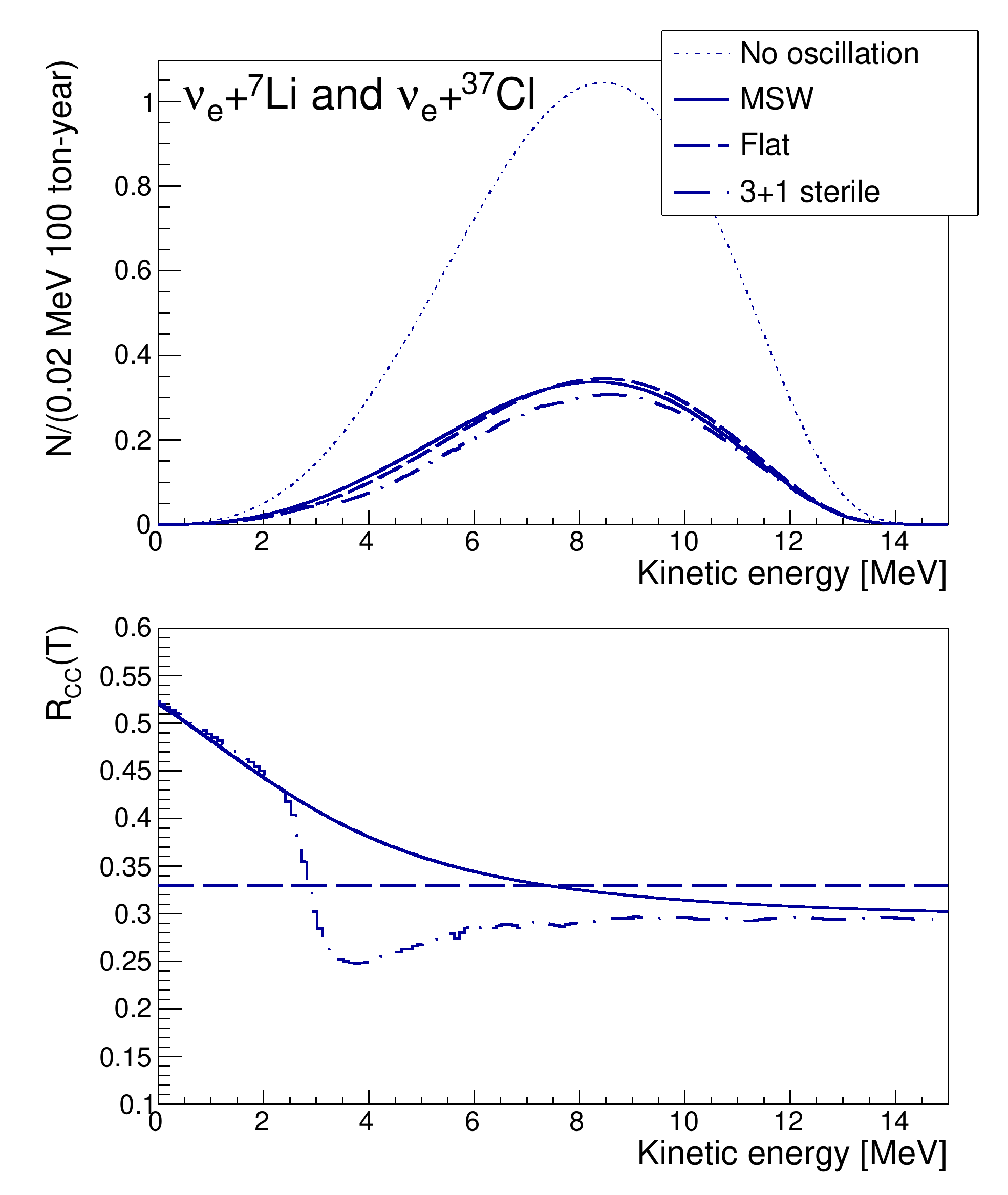} % Here is how to import EPS art
\caption{\label{fig:B8Fig2}
The final state effective kinetic energy spectra of the LiCC and ClCC signal events of $^8$B neutrinos
with different oscillation configurations (upper panel) and the ratios of the oscillated spectra to the undisturbed spectrum (lower panel).
For the sterile neutrino mixing, $\alpha$ = 0.021 and $\Delta m^2_{01}$ = $1.56\times10^{-5}$~eV$^2$.
The exposure is set to 100-ton LiCl solution $\times$ 1 data-taking year, and
the LiCl concentration in the water solution is assumed to be 74.5~g/100~g water.}
\end{figure}

The neutrino event rates and spectra of LiCC and ClCC are estimated according to the cross-sections and
the molarities of $\rm{{}^{7}Li}$ and $\rm{{}^{37}Cl}$ in the LiCl solution.
The event rate, $N_{\rm{LiCC}}$, as a function of the kinetic energy (defined in Eq.~\ref{eq:EdetLi}) is calculated as
\begin{equation}
\begin{split}
N_{\rm{LiCC}}(T)=t N_{\rm{Li}} \sum_i^{\rm{levels}} \Phi_{\nu}(E_{\nu}) \sigma_{\rm{Li-i}}(E_{\nu},T) P_{ee}(E_{\nu}),
\end{split}
\label{eq:NLiCC}
\end{equation}
where $\sigma_{\rm{Li-i}}(E_{\nu},T)$ is calculated in Eq.~\ref{eq:EdetLi}, \ref{eq:Sigma}, \ref{eq:Sigma0}, and \ref{eq:omega}, including all final state levels of $\rm{{}^{7}Be}$,
$t$ is the data-taking time, and $N_{\rm{Li}}$ is the number of target $\rm{{}^{7}Li}$ per unit LiCl water solution as described in Sec.~\ref{sec:detectorproposal}.
Similarly, we obtain the event rate and spectrum, $N_{\rm{ClCC}}(T)$, for the ClCC process.
The energy spectrum of all CC events on nuclei is
\begin{equation}
\begin{split}
N_{\rm{CC}}(T)=N_{\rm{LiCC}}(T)+N_{\rm{ClCC}}(T).
\end{split}
\label{eq:NCC}
\end{equation}
Both the oscillated and undistorted $T$ spectra of $^8$B neutrino CC events are shown in Fig.~\ref{fig:B8Fig2}.
The integrated LiCC, ClCC, and all CC rates with undistorted $^8$B neutrino spectrum, oscillated spectrum, and oscillated spectrum plus a $T>5$~MeV cut are calculated and tabulated in Tab.~\ref{tab:EventRate}. The ClCC event rate is 7\% of the LiCC process.

\begin{figure}[!htbp]
\includegraphics[width=0.45\textwidth]{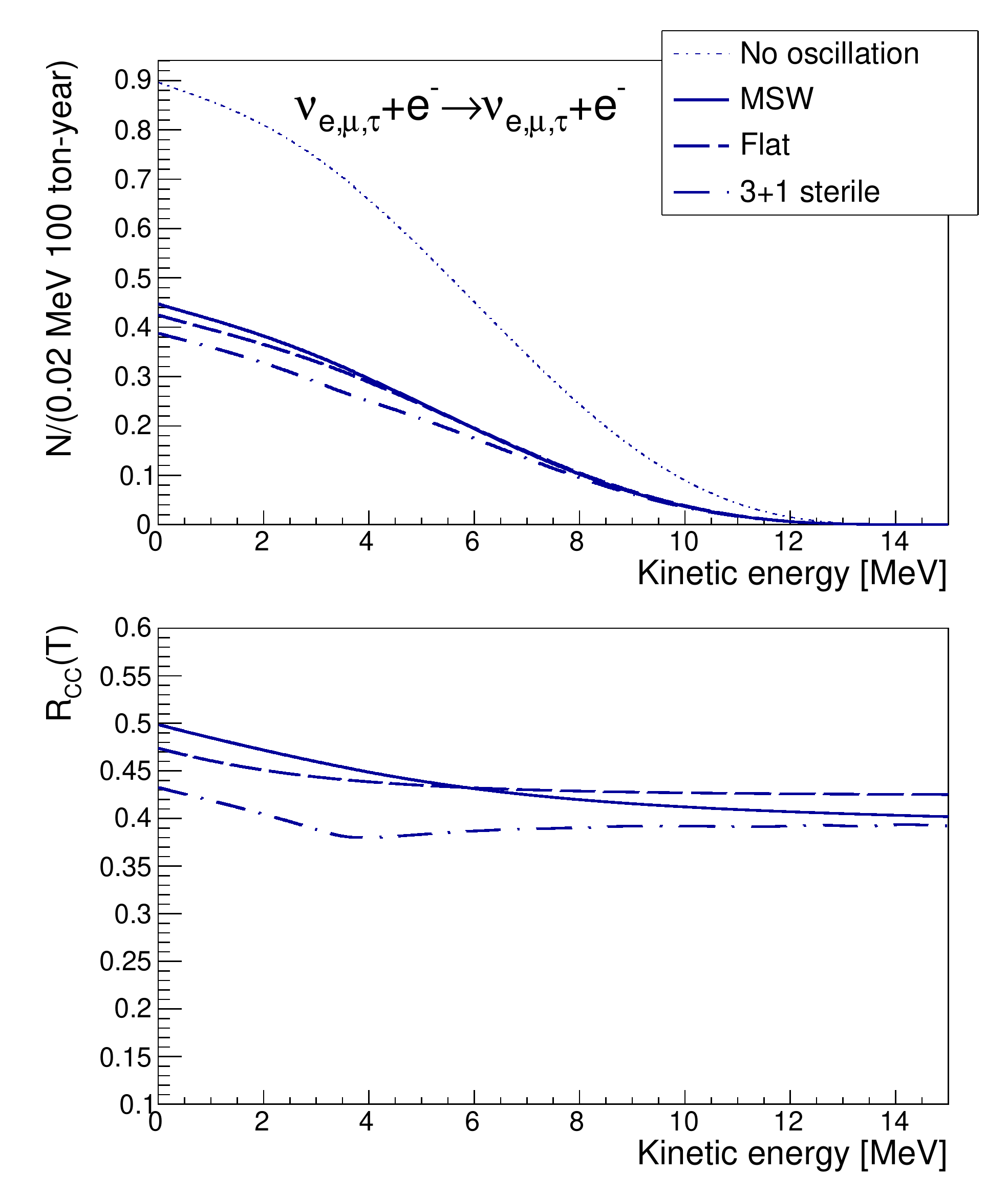} % Here is how to import EPS art
\caption{\label{fig:B8Fig3}
The recoil electron kinetic energy spectra of the Elas signal events of $^8$B neutrinos
with different oscillation configurations (upper panel) and the ratios of the oscillated spectra to the undisturbed spectrum (lower panel).
For the sterile neutrino mixing, $\alpha$ = 0.021 and $\Delta m^2_{01}$ = $1.56\times10^{-5}$~eV$^2$.
The exposure is set to 100-ton LiCl solution $\times$ 1~data-taking year, and
the LiCl concentration in the water solution is assumed to be 74.5~g/100~g water.}
\end{figure}

The surviving $\nu_e$, $\nu_\mu$, and $\nu_\tau$ all scatter on electrons.
The kinetic energy spectrum, $N_{\rm{Elas}}(T)$, of recoil electrons contains all three contributions, and
it is
\begin{equation}
\begin{split}
N_{\rm{El}}(T)=&t N_e \int dE_{\nu} \Phi_{\nu}(E_{\nu}) \{ \sigma_e(E_{\nu},T) P_{ee}(E_{\nu})\\
        &+ \sigma_{\mu,\tau}(E_{\nu},T) \left[1-P_{ee}(E_{\nu})\right]  \},
\end{split}
\label{eq:NElas}
\end{equation}
where $\sigma_e(E_{\nu},T)$ and $\sigma_{\mu,\tau}(E_{\nu},T)$ are the differential scattering cross sections
as a function of electron kinetic energy for $\nu_{e}$ and $\nu_{\mu,\tau}$~\cite{Elastic}, respectively, and
$N_e$ is the total number of target electrons per unit LiCl water solution as described in Sec.~\ref{sec:detectorproposal}.
Both the oscillated and undistorted $T$ spectra of the $^8$B neutrinos of the Elas process are shown in Fig.~\ref{fig:B8Fig3}
The integrated Elas event rates with the undistorted $^8$B neutrino spectrum, oscillated spectrum, and oscillated spectrum plus a $T>5$~MeV cut are calculated and tabulated in Tab.~\ref{tab:EventRate}.
The ratio of the CC event rate to that of Elas events is 108:124 for the oscillated spectrum,
and it is further enhanced to 94:35 with the $T>5$~MeV cut.
In Fig.~\ref{fig:angular}, correspondingly, the ratio of the number of CC events to the number of Elas events is set according to 94:35.

In summary, as seen in Tab.~\ref{tab:EventRate}, the total event rates of CC, mainly LiCC, and Elas are comparable in the proposed LiCl water solution. With the $T>5$~MeV cut, the CC events are dominant.

\subsection{Upturn effect}
\label{sec:upturn}
In the following discussion, we compare the signal strengths of the upturn effect of the CC and Elas processes.

In Fig.~\ref{fig:B8Fig1} of the $P_{ee}(E_{\nu})$ of the MSW solution,
a clear upturn can be seen from the high energy to low energy.
To obtain a specific quantitative size of upturn, the relative difference between the survival probability at 4.862~MeV and 10,862~MeV is calculated as a figure of merit, and the result is
\begin{eqnarray}
\label{eq:PeeRatio}
\frac{P_{ee}(4.862~{\rm{MeV}}) - P_{ee}(10.862~{\rm{MeV}})}{P_{ee}(10.862~{\rm{MeV}})} = 23\%.
\end{eqnarray}
Subtracting the $\rm{{}^{7}Li}$ interaction threshold (Eq.~\ref{eq:EdetLi}), they correspond to 4 and 10~MeV for the LiCC final state effective kinetic energy $T$.

As a reference, a flat survival probability $P_{ee}$ is plotted in Fig.~\ref{fig:B8Fig1}, in which $P_{ee}$ is set to a constant of 0.33,
and $P_{e\mu}+P_{e\tau}$ = 0.67.

With the CC interactions on $\rm{{}^{7}Li}$ or $\rm{{}^{37}Cl}$, the final state effective kinetic energy spectrum
is calculated with Eq.~\ref{eq:NCC} and shown in Fig.~\ref{fig:B8Fig2}.
We define $R_{\rm{CC}}(T)$ as
\begin{eqnarray}
\label{eq:RatioLi}
R_{\rm{CC}}(T)=\frac{N_{\rm{CC}}(T)~|~\rm{Osci}}{N_{\rm{CC}}(T)~|~\rm{No~Osci}},
\end{eqnarray}
which is the ratio of the kinetic energy spectra under the oscillation condition (Osci) to no oscillation (No Osci).
The spectrum of $R_{\rm{CC}}(T)$ is plotted in the lower panel of Fig.~\ref{fig:B8Fig2}.
The relative difference between $R_{\rm{CC}}(4~{\rm{MeV}})$ and $R_{\rm{CC}}(10~{\rm{MeV}})$ for the MSW oscillation study is
\begin{eqnarray}
\label{eq:RCCRatio}
\rm{MSW:}~\frac{R_{\rm{CC}}(4~{\rm{MeV}}) - R_{\rm{CC}}(10~{\rm{MeV}})}{R_{\rm{CC}}(10~{\rm{MeV}})} = 23\%.
\end{eqnarray}
This CC channel result has a consistent signal strength with the original result in Eq.~\ref{eq:PeeRatio}.
For the flat $P_{ee}$ configuration, there is no interesting feature in the $R_{\rm{CC}}(T)$ spectrum, as plotted in Fig.~\ref{fig:B8Fig2}.

For the Elas process, the kinetic energy spectrum of the recoil electron is calculated according to Eq.~\ref{eq:NElas}
and shown in Fig.~\ref{fig:B8Fig3}.
A similar ratio of $R_{\rm{El}}(T)$ is defined as
\begin{eqnarray}
\label{eq:RatioEl}
R_{\rm{El}}(T)=\frac{N_{\rm{El}}(T)~|~\rm{Osci}}  {N_{\rm{El}}(T)~|~\rm{No~Osci}},
\end{eqnarray}
which is the ratio of the oscillated kinetic energy spectrum to the undistorted spectrum.
The spectrum of $R_{\rm{El}}(T)$ is plotted in the lower panel of Fig.~\ref{fig:B8Fig3}.
The relative difference between $R_{\rm{El}}(4~{\rm{MeV}})$ and $R_{\rm{El}}(10~{\rm{MeV}})$ for the MSW oscillation study is
\begin{eqnarray}
\label{eq:RatioElMSW}
{\rm{MSW}:}~\frac{R_{\rm{El}}(4~{\rm{MeV}}) - R_{\rm{El}}(10~{\rm{MeV}})}{R_{\rm{El}}(10~{\rm{MeV}})} = 9.6\%.
\end{eqnarray}
However, we notice that in the Elas process, even for the flat survival probability, there is a minor upturn in the ratio plot,
which is shown in the lower panel of Fig.~\ref{fig:B8Fig3}.
The relative difference
\begin{eqnarray}
\label{eq:RatioElFlat}
{\rm{flat}:}~\frac{R_{\rm{El}}(4~{\rm{MeV}}) - R_{\rm{El}}(10~{\rm{MeV}})}{R_{\rm{El}}(10~{\rm{MeV}})} = 2.7\%.
\end{eqnarray}
This is caused by the difference in the differential cross-section in $\nu_e$-e and $\nu_{\mu,\tau}$-e.
An extra $\nu_{\mu,\tau}$ contribution appears to the low energy part of the recoil electron spectrum.
Therefore, the net signal strength with the Elas process is only 6.9\%, which is much smaller than the 23\% of Eq.~\ref{eq:PeeRatio} and Eq.~\ref{eq:RCCRatio}.

In summary, in such a LiCl detector, as described in section~\ref{sec:detector},
the strength of the upturn signal of the CC processes on LiCl is much larger than that of the Elas process.

\subsection{Sterile neutrino}
\label{sec:sterile}
The neutrino transition probability $P_{ee}(E_{\nu})$, $P_{e\mu}(E_{\nu})$, and $P_{e\tau}(E_{\nu})$ calculation with the (3+1) situation~\cite{Sterile1, Sterile2} is done in the following way.

We adopt the parameter convention of sterile neutrinos in~\cite{Sterile1, Sterile2}.
One parameter, mixing angle $\alpha$, describes the mixing between the sterile neutrino, $\nu_s$, and active neutrinos, and
the other parameter, mass squared difference $\Delta m^2_{01}$, is the mass squared difference between $\nu_0$ and $\nu_1$, in which
$\nu_0$ is introduced along with $\nu_s$.
With a small mixing angle $\alpha$, $\nu_s$ mixes weakly with active neutrinos, and $\nu_s$ almost coincides with $\nu_0$.

The nonadiabatic situation must be considered for the propagation of (3+1) neutrinos in the Sun for some parameter settings of $\alpha$ and $\Delta m^2_{01}$.
The $^8$B neutrino generation zone $r<0.135 R_{\odot}$~\cite{NumberDensity} is divided into $60\times60\times60$ ($x\times y\times z$) small cells, and $^8$B neutrinos are generated according to the probability density function~\cite{NumberDensity}.
We use two different numerical methods to calculate the flavor transition probability $P_{ee}(E_{\nu})$, $P_{e\mu}(E_{\nu})$, and $P_{e\tau}(E_{\nu})$.
One is the multislab method.
The neutrino outgoing path is divided into many slabs with a step size of $R_{\odot}/1000$.
Besides the local number density of electrons, the local number density of neutrons~\cite{NumberDensity} is also considered for the important neutral current process~\cite{Sterile1, Sterile2}.
Each slab is assumed to have a uniform material with constant number densities.
Flavor and mass eigenstate transfer occurs at each slab interface.
The other method is a 4th-order Runge-Kutta method, which is used to solve the propagation differential equations with $10^6$ steps.
The probabilities $P_{ee}(E_{\nu})$, $P_{e\mu}(E_{\nu})$, and $P_{e\tau}(E_{\nu})$ calculated by the two methods are in good agreement in our test, as in~\cite{thesis}.
The multislab method is easier and faster to obtain a stable solution for this calculation.
No Earth matter effect is considered since the difference introduced by the Earth effect is not significant for this study~\cite{SNOThesis}.

Taking one set of sterile neutrino mixing parameters ($\alpha$ = 0.021 and $\Delta m^2_{01}$ = $-1.56\times10^{-5}$~eV$^2$) as an example,
the oscillated neutrino spectrum and the $\nu_e$ survival probability $P_{ee}(E_{\nu})$ are shown in Fig.~\ref{fig:B8Fig1}.
The kinetic energy spectrum of the CC processes and its ratio to the undistorted spectrum are both shown in Fig.~\ref{fig:B8Fig2}.
The recoil electron kinetic energy spectrum of the Elas process and its ratio to the undistorted spectrum are both shown in Fig.~\ref{fig:B8Fig3}.

With these comparisons, we see that
the rich structure information in Fig.~\ref{fig:B8Fig1} is reserved in the CC process, as shown in Fig.~\ref{fig:B8Fig2};
however, it is almost smeared out in Fig.~\ref{fig:B8Fig3} with the Elas process.
To distinguish the sterile neutrinos with the CC channels, fewer signal statistics are needed.

\subsection{Earth matter effect}
\label{sec:earth}
In this section, we follow the three-active-neutrino oscillation calculation in Sec.~\ref{tab:EventRate}
and then continue with the oscillation calculation of the Earth matter effect.
A multilayer Earth model~\cite{Earth} is adopted, and a multislab numerical calculation (Sec.~\ref{sec:sterile}) is implemented.

\begin{figure}[!htbp]
\includegraphics[width=0.48\textwidth]{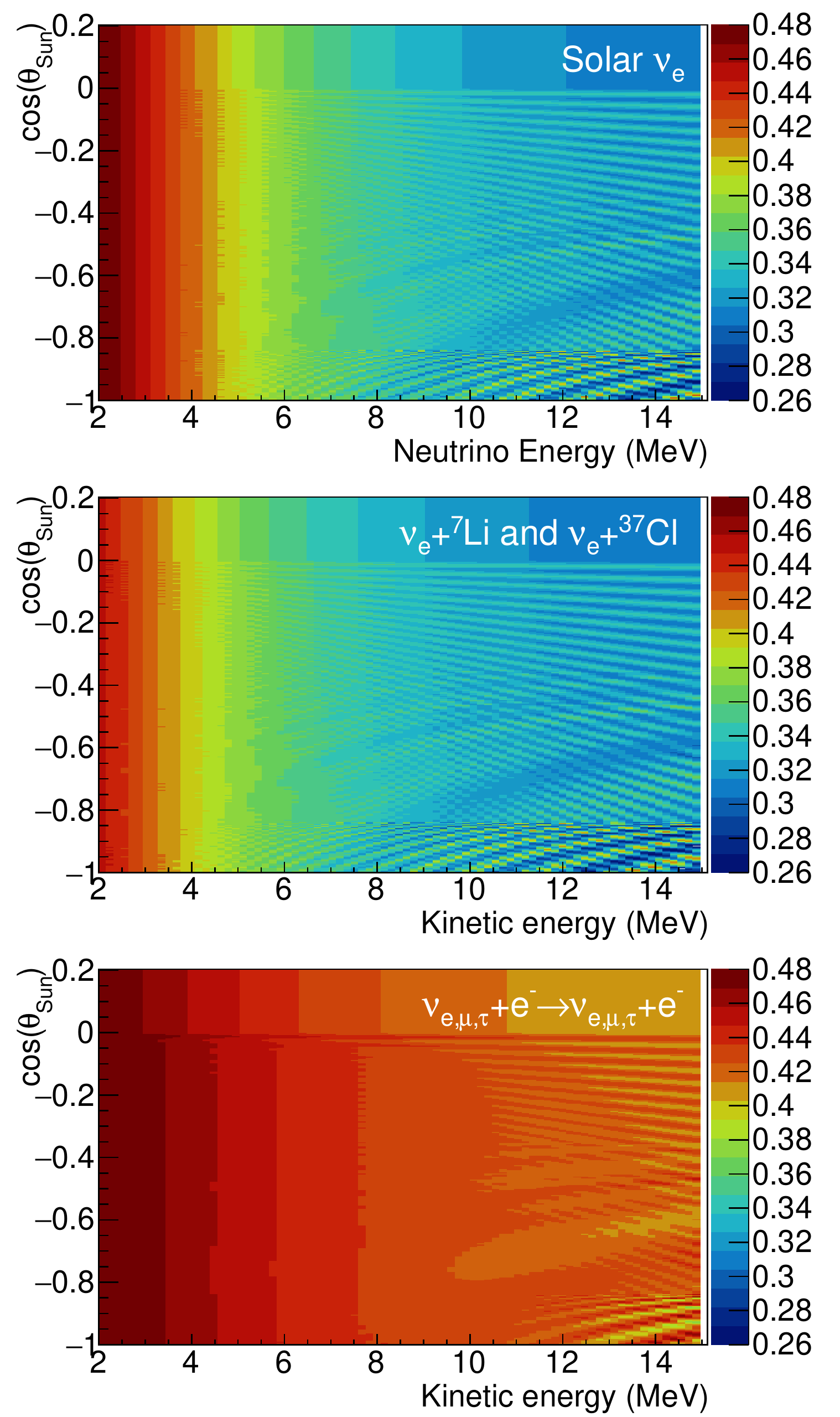}
\caption{\label{fig:Oscillogram}
Electron neutrino survival probability $P_{ee}$ as a function of the neutrino energy and the cosine of the solar angle (upper).
For the neutrino going through the Earth, $\cos(\theta_{\rm{Sun}})$ is negative.
Ratio of the oscillated LiCl charged-current event rate to the undistorted LiCl charged-current event rate, i.e., $R_{\rm{CC}}$, in Eq.~\ref{eq:RatioLi}
as a function of the kinetic energy of the final state particles and the cosine of the solar angle (middle).
Ratio of the oscillated Elas event rate to the undistorted Elas event rate, i.e., $R_{\rm{El}}$ in Eq.~\ref{eq:RatioEl}
as a function of the recoil electron kinetic energy and the cosine of the solar angle (lower).}
\end{figure}

In the upper panel of Fig.~\ref{fig:Oscillogram}, we show the $\nu_e$ survival probability $P_{ee}$
as a function of the neutrino energy, $E_{\nu}$, and the cosine of the solar angle, $\cos(\theta_{\rm{Sun}})$.
For the neutrino going through the Earth, $\cos(\theta_{\rm{Sun}})$ is negative.
The pattern can be understood as in~\cite{tomograph}.

With the CC processes on LiCl, the kinetic energy spectrum is obtained for each solar angle as done with Eq.~\ref{eq:NCC}.
In the middle panel of Fig.~\ref{fig:Oscillogram}, we show the $R_{\rm{CC}}$ (defined in Eq.~\ref{eq:RatioLi})
as a function of the kinetic energy, $T$, and $\cos(\theta_{\rm{Sun}})$.

The lower panel of Fig.~\ref{fig:Oscillogram} is the plot for $R_{\rm{El}}$ (defined in Eq.~\ref{eq:RatioEl}) for the Elas process.

First, the original rich pattern in the neutrino plot (upper panel of Fig.~\ref{fig:Oscillogram}) is well repeated in the LiCl CC plot (middle panel) but is almost smeared out in the Elas plot (lower panel).
Second, the structures are mostly seen in the 4-12~MeV region of the neutrino plot (upper and middle panels),
where the CC kinetic energy spectrum has the most statistics, as seen in Fig.~\ref{fig:B8Fig2}.
There are some residual structures in the Elas plot higher than 10~MeV (lower panel),
but the statistics of the Elas signals is low, as seen in Fig.~\ref{fig:B8Fig3}.
In conclusion, the CC process is most helpful in distinguishing the Earth matter effect.

\section{Sensitivity study for upturn and sterile neutrino}
\label{sec:sensitivity}
In this section, more realistic detector effects and signal selection criteria are considered.
The sensitivity of the upturn study and sterile neutrino search is presented.
The Earth matter effect study requires a different detector setup, such as a higher energy resolution and higher signal statistics,
and the required detector configuration and sensitivity are given in a separate paper.

\subsection{Detector effects and signal selection criteria}
\label{sec:DetectorEff}
With the detector proposal and expected property described in Sec.~\ref{sec:detectorproposal}, more realistic predictions are made, and
based on the predictions, many random samples are generated for the following sensitivity studies.

A gauss energy smearing with an energy resolution of 20~PE/MeV, $R(T, E_{\rm{rec}})$, is applied to the kinetic energy spectrum
of $N_{\rm{CC}}(T)$ (Eq.~\ref{eq:NCC}) and $N_{\rm{El}}(T)$ (Eq.~\ref{eq:NElas}).
\begin{equation}
\begin{split}
N_{\rm{CC}}(E_{\rm{rec}}) = N_{\rm{CC}}(T) \otimes R(T, E_{\rm{rec}}),\\
N_{\rm{El}}(E_{\rm{rec}}) = N_{\rm{El}}(T) \otimes R(T, E_{\rm{rec}}),
\end{split}
\label{eq:ERec}
\end{equation}
where $E_{\rm{rec}}$ is the reconstructed energy, and
the corresponding energy spectra $N_{\rm{CC}}(E_{\rm{rec}})$ and $N_{\rm{El}}(E_{\rm{rec}})$ are obtained for the CC and Elas processes, respectively.
For the limited data-taking time and target mass and a binned fitting later,
each $E_{\rm{rec}}$ spectrum is divided into several 1~MeV bins, and
the bin content in each bin is denoted with $N_{\rm{CC},i}$ or $N_{\rm{El},i}$, where $i$ is the bin number.

The CC and Elas events are the background to each other.
As discussed in Sec.~\ref{sec:detectorproposal}, we expect that there is no other background with an $E_{\rm{rec}}>$5~MeV cut or a more aggressive $E_{\rm{rec}}>$4~MeV cut.

\begin{figure}[]
\includegraphics[width=0.48\textwidth]{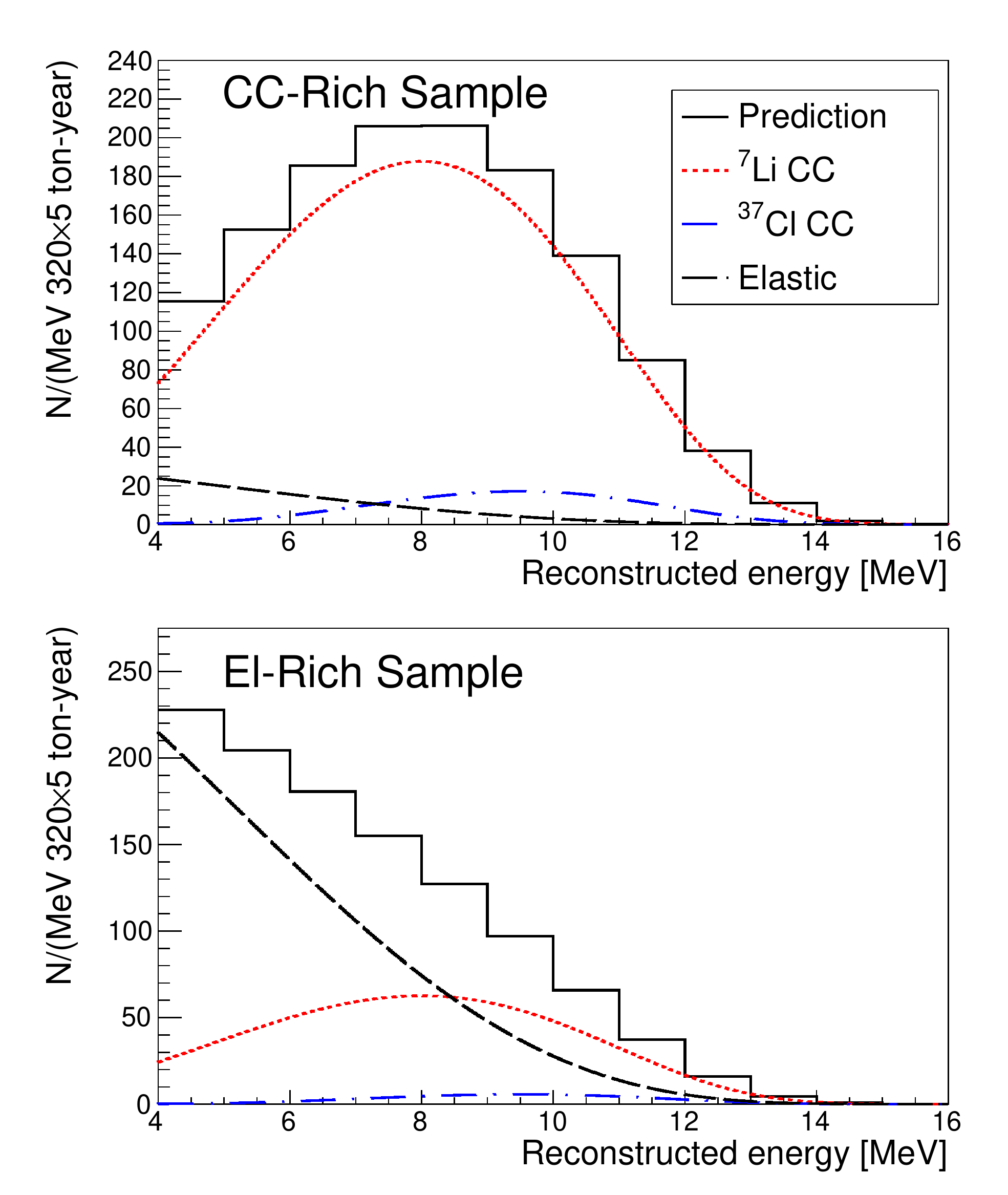}
\caption{\label{fig:Li7CC-rich}
Predicted CC\text{-}rich (upper) and El\text{-}rich (lower) samples as a function of $E_{\rm{rec}}$ are shown. The contained LiCC, ClCC, and Elas components are also overlaid.
The exposure is set to 320-ton LiCl solution $\times$ 10 data-taking years, and
the LiCl concentration in the water solution is assumed to be 74.5 g/100 g water.
The detector resolution and signal selection information can be seen in Sec.~\ref{sec:DetectorEff}.}
\end{figure}

The Elas and CC signals must be separated with a reconstructed solar angle cut (see Fig.~\ref{fig:angular}).
Applying a solar angle cut at 60~degrees, i.e., $\cos(\theta_{\rm{Sun}})=0.5$,
a CC\text{-}rich sample with 10\% of the Elas events and 75\% of the CC events and
an El\text{-}rich sample with 90\% of the Elas events and 25\% of the CC events are obtained.
They are represented by
\begin{equation}
\begin{split}
N_{{\rm{CC\text{-}rich}},i} = N_{{\rm{CC}},i}75\% + N_{{\rm{El}},i}10\%,\\
N_{{\rm{El\text{-}rich}},i} = N_{{\rm{CC}},i}25\% + N_{{\rm{El}},i}90\%.
\end{split}
\label{eq:CCElRich}
\end{equation}
The predicted $N_{\rm{CC\text{-}rich},i}$ and $N_{\rm{El\text{-}rich},i}$ are shown in Fig.~\ref{fig:Li7CC-rich}, and
the LiCC, ClCC, and Elas components are also shown.

\subsection{Upturn effect}
\label{sec:UpturnSens}
Poisson random sampling is performed according to the predictions in Eq.~\ref{eq:CCElRich} for an exposure of 320$\times$5 ton-year.
A total of 1000 random samples are generated, and for each sample, there are two sub datasets,
$D_{{\rm{CC\text{-}rich}},i}$ and $D_{{\rm{El\text{-}rich}},i}$, for the CC\text{-}rich and El\text{-}rich, respectively.

Each random sample is fitted with the following $\chi^2$
\begin{align}
\begin{split}
\chi^2 = &\sum^{}_i (D_{{\rm{CC\text{-}rich}},i} - P_{{\rm{CC\text{-}rich}},i})^2/D_{{\rm{CC\text{-}rich}},i} \\
         &+\sum^{}_i (D_{{\rm{El\text{-}rich}},i} - P_{{\rm{El\text{-}rich}},i})^2/E_{{\rm{El\text{-}rich}},i} \\
         &+{\rm{Pull}},
\label{eq:chi2}
\end{split}
\end{align}
in which the predictions are $P_{{\rm{CC\text{-}rich}},i}$ and $P_{{\rm{El\text{-}rich}},i}$ for the CC-rich and El-rich samples, respectively.
With respect to Eq.~\ref{eq:CCElRich}, they further consider the detector systematic uncertainties as follows for the fit.
\begin{align}
\begin{split}
\label{eq:P}
P_{{\rm{CC\text{-}rich}},i} = &[N_{{\rm{CC}},i} 75\%(1+\eta_{\sigma}) + N_{{\rm{El}},i} 10\%(1+\eta_{\epsilon})] \\
                       &\times(1+\eta_{\rm{Norm}}), \\
P_{{\rm{El\text{-}rich}},i} = &[N_{{\rm{CC}},i} 25\%(1+\eta_{\sigma}) + N_{{\rm{El}},i} (1-10\%(1+\eta_{\epsilon}))] \\
                       &\times(1+\eta_{\rm{Norm}}),  \\
\end{split}
\end{align}
where the LiCC cross-section uncertainty, $\eta_{\sigma}$, the selection efficiency uncertainty of the Elas process, $\eta_{\epsilon}$, and the normalization uncertainty, $\eta_{Norm}$ are taken into account.
The $^8$B neutrino spectrum theoretical error in the energy region of interest~\cite{B8spectrum4} is very small compared to the expected statistical error, and the theoretical spectrum uncertainty is ignored in this study.
The uncertainty of the LiCC cross-section is assumed to be 2\%, as suggested in Sec.~\ref{sec:cross-section}.
The Elas process selection efficiency uncertainty requires a dedicated detector calibration and is assigned as 10\%.
The normalization uncertainty includes the impact of the high-low metallicity ambiguity~\cite{LowMetallicity} and fiducial mass and is assigned as 10\%.
The fit basically only considers the shape of the observed spectrum.
Correspondingly, the pull term is
\begin{align}
\begin{split}
{\rm{Pull}} = (\eta_{\sigma}/2\%)^2 + (\eta_{\epsilon}/10\%)^2 + (\eta_{\rm{Norm}}/10\%)^2.
\label{eq:Pull}
\end{split}
\end{align}

\begin{figure}[]
\includegraphics[width=0.48\textwidth]{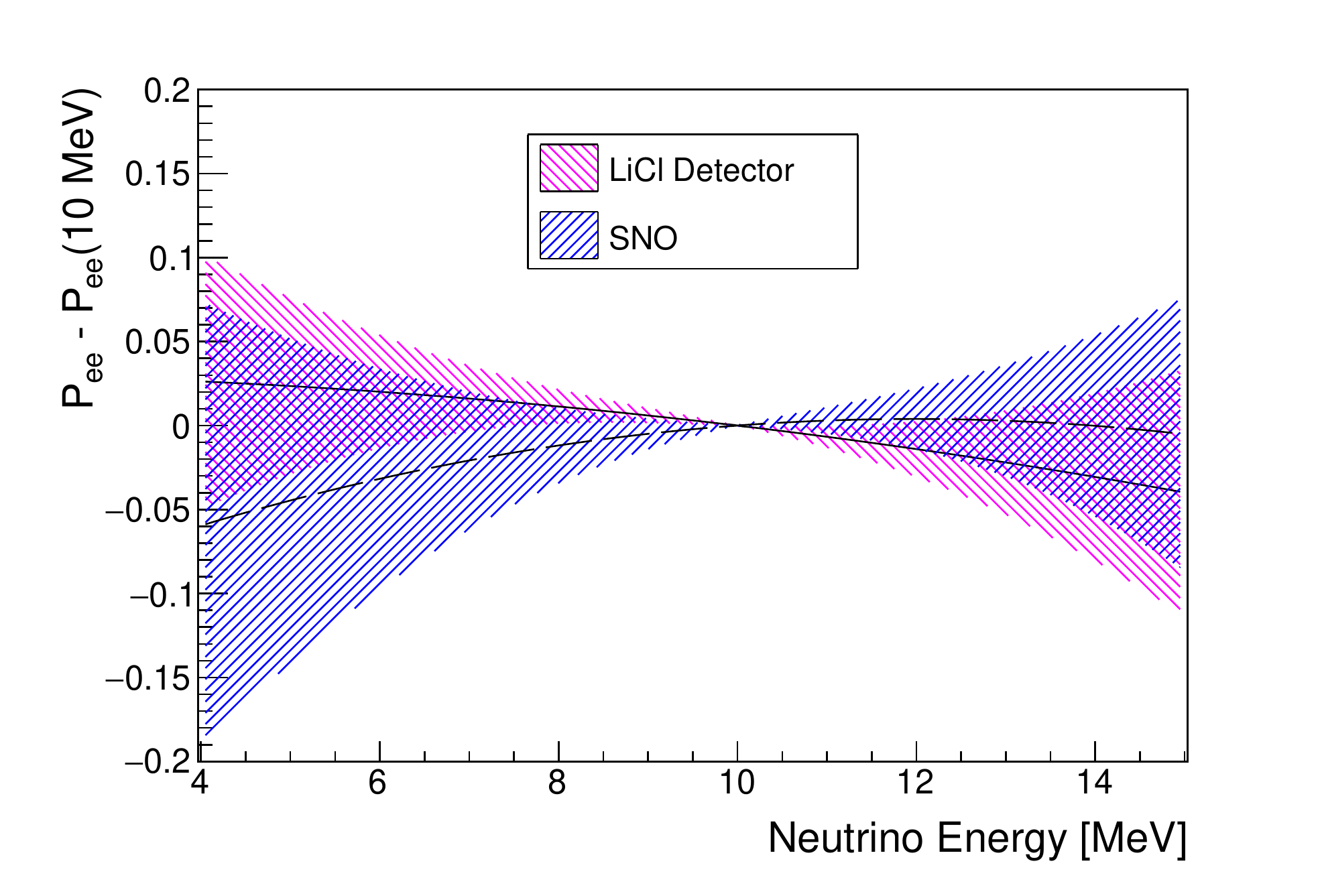}
\includegraphics[width=0.48\textwidth]{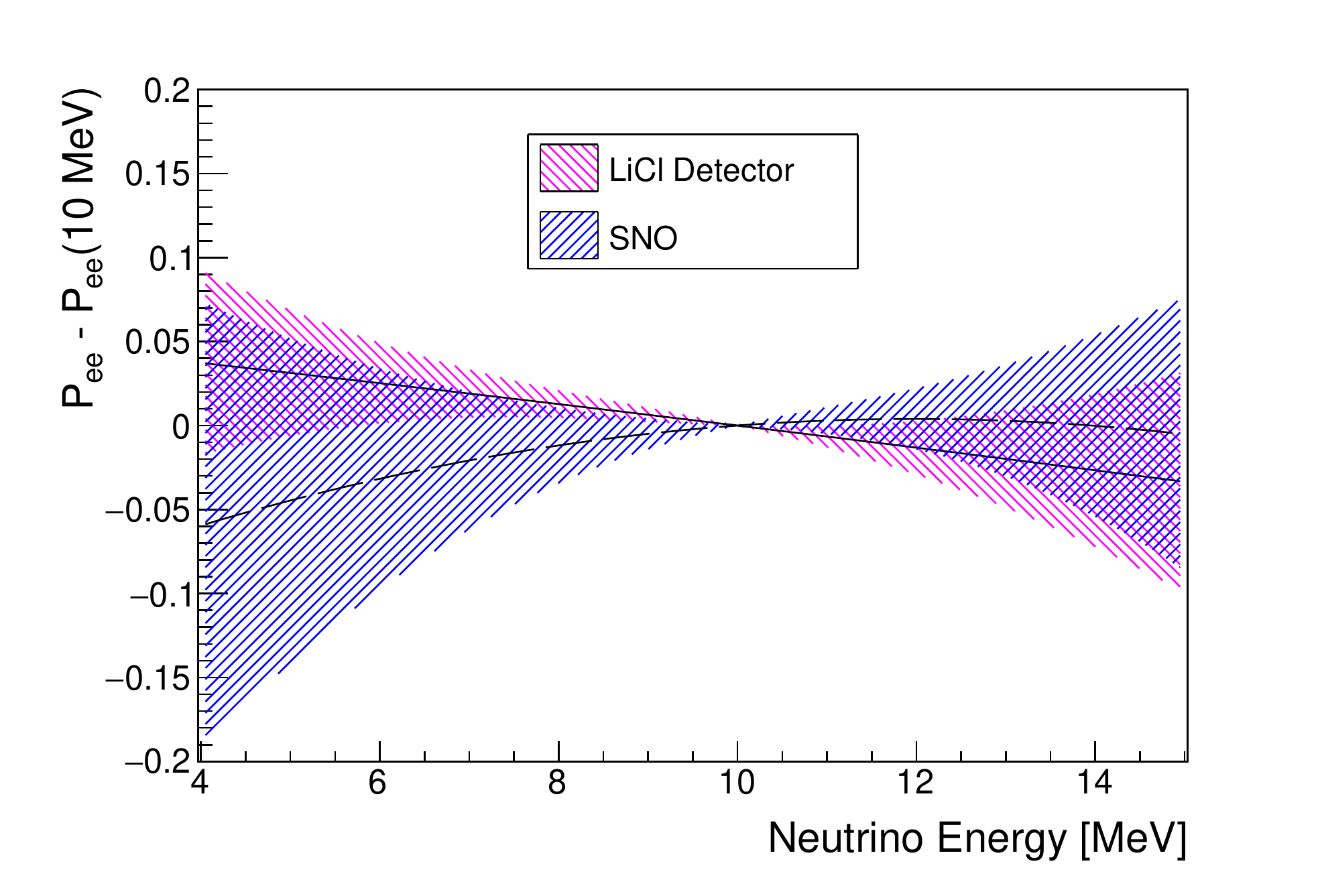}
\caption{\label{fig:UpturnSens}
The best fit and rms spread of $c_1(E_{\nu}-10)+c_2(E_{\nu}-10)^2$ of $P_{ee}(E_{\nu})$ in Eq.~\ref{eq:Osci2nd}.
The upper panel is the fit result with a 5~MeV cut and the lower panel is with a 4~MeV cut.
They present the evolution trend of $\nu_e$ survival probability as a function of $E_{\nu}$ in the upturn sensitive region.
The result from the SNO experiment~\cite{SNO} is overlaid for comparison.
The LiCl water solution detector has an exposure of 320$\times$5 ton-year,
where the LiCl concentration is assumed to be 74.5 g/100 g water.}
\end{figure}

For the upturn study, we adopt two kinds of simplified $P_{ee}(E_{\nu})$ for Eq.~\ref{eq:P}.
The first one is a quadratic function of $E_{\nu}$ as used in the SNO~\cite{SNO} and Super-Kamiokande~\cite{SK} experiments.
Using the 10~MeV survival probability as the reference point, the survival probability is
\begin{align}
\begin{split}
P_{ee}(E_{\nu})    &= c_0+c_1(E_{\nu}-10)+c_2(E_{\nu}-10)^2.
\label{eq:Osci2nd}
\end{split}
\end{align}
The term $c_0$ is the best fit of the average survival probability at 10~MeV, which is not the topic of this paper.
To focus on the upturn issue,
the best fit and rms spread of $c_1(E_{\nu}-10)+c_2(E_{\nu}-10)^2$ for the LiCl water solution detector with the exposure of 320$\times$5 ton-year are shown in Fig.~\ref{fig:UpturnSens}, where both the 5~MeV cut and 4~MeV cut results are analyzed.
The result from the SNO experiment is also overlaid for comparison.
A significant improvement with the compact LiCl detector can be expected.

\begin{figure}
\includegraphics[width=0.48\textwidth]{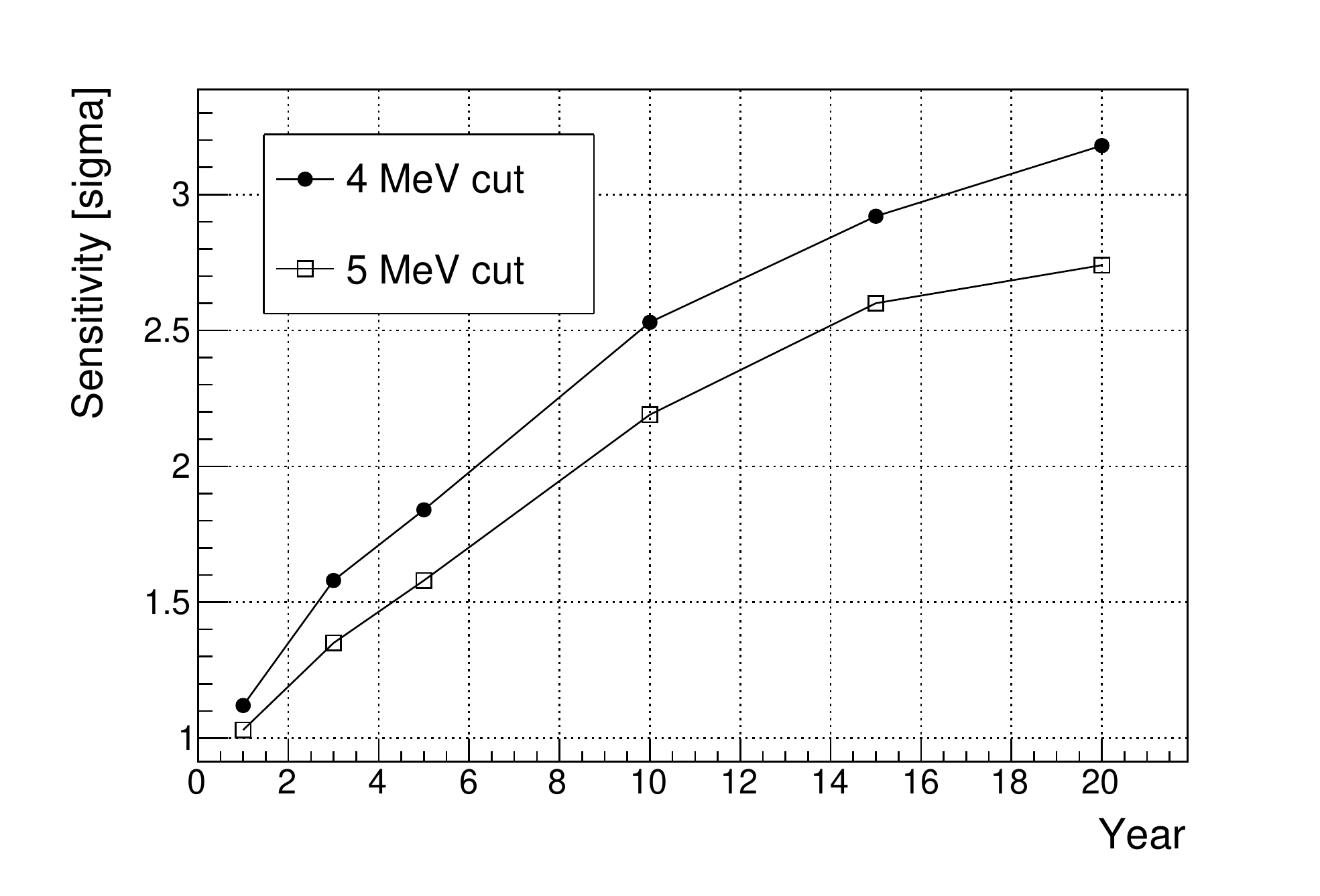} % Here is how to import EPS art
\caption{
\label{fig:UpturnScan}
The sensitivity of rejecting no upturn effect versus the number of data-taking years for a 320-ton LiCl water solution.
More detector resolution and signal selection information can be seen in Sec.~\ref{sec:DetectorEff}.}
\end{figure}

The sensitivity of observing the upturn is scanned with exposures of 320 tons$\times$1, 3, 5, 10, 15, and 20 years.
One thousand data samples are generated with each exposure setting.
Because of the joint contribution of $c_1$ and $c_2$, the upward or downward slope is difficult to quantify.
The $P_{ee}(E_{\nu})$ in Eq.~\ref{eq:Osci2nd} is replaced by a simple linear function as below
\begin{align}
\begin{split}
P_{ee}(E_{\nu})    &= c_0+c_1(E_{\nu}-10).
\label{eq:Osci1st}
\end{split}
\end{align}
If a statistically significant negative fit result of $c_1$ is obtained, an upturn is observed.
On the contrary, a zero or positive fit result of $c_1$ corresponds to no upturn effect, i.e., the null assumption.
The 1000 $c_1$ fit results of each exposure setting are plotted.
They are gauss distributed, and the mean $\mu_{c1}$ and resolution $\sigma_{c1}$ are obtained.
For our proposed detector setup and exposures, $\mu_{c1}$ is always negative, i.e., the upturn is favored.
Therefore, we define $\mu_{c1}/\sigma_{c1}$ as the sensitivity for rejecting the null assumption.
The sensitivity versus the exposures is shown in Fig.~\ref{fig:UpturnScan},
and the results with the 5~MeV and 4~MeV cuts are both shown.

In summary, we expect an improvement in studying the upturn effect with the LiCl water solution detector.

\subsection{Sterile neutrino}
\label{sec:SterileSens}
\begin{figure}[]
\includegraphics[width=0.48\textwidth]{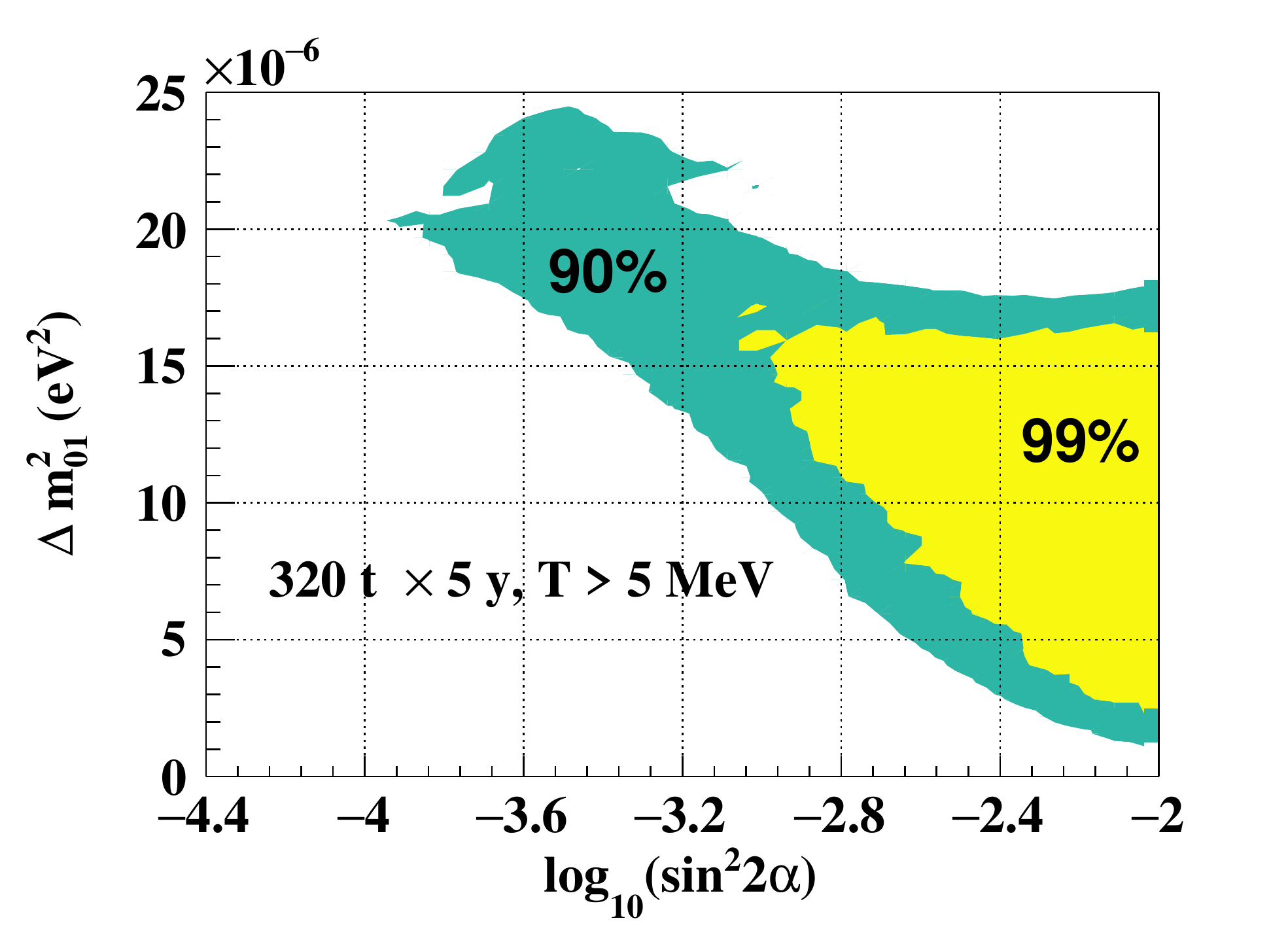}
\caption{
Exclusion sensitivity of sterile neutrinos using a LiCl detector with an exposure of 320$\times$5 ton-year.
More detector resolution and signal selection information can be seen in Sec.~\ref{sec:DetectorEff}.}
\label{fig:SterileSensi}
\end{figure}

We use the Feldman-Cousin method~\cite{FC} to determine the exclusion sensitivity~\cite{Exclusion} of a detector with the exposure of 320$\times$5 ton-year.
The $\alpha$ and $\Delta m^2_{01}$ parameter space, $\log_{10}(\sin^22\alpha)\in$ $[-5$, $-2]$ and $\Delta m^2_{01}\in$ $[0$, $25\times10^{-6}]~eV^2$, is split into 40$\times$40 grids.
For each grid, 1000 statistically random samples are generated according to the exposure configuration.
The $P_{ee}(E_{\nu})$ calculation procedure is described in Sec.~\ref{sec:sterile}, and
the detector effect is added as in Eq.~\ref{eq:ERec} and \ref{eq:CCElRich} of Sec.~\ref{sec:DetectorEff}.
With the $\chi^2$ definition in Eq.~\ref{eq:chi2}, \ref{eq:P} and \ref{eq:Pull}, the exclusion sensitivity is estimated.
Figure.~\ref{fig:SterileSensi} shows the sensitivity contours with a 5~MeV cut.
Besides the statistical power, the proposed detector and settings are not sensitive to some sterile parameter regions due to the 5~MeV cut and the systematic uncertainties.

\section{Conclusion}
\label{sec:conclude}
In this work, we study MeV neutrino detection in LiCl water solution.
We reevaluate the $\nu_e$ charged-current interaction cross-section on $\rm{{}^{7}Li}$ with new B(GT) experimental inputs.
The total CC interaction cross-section weighted by the solar $^8$B electron neutrino spectrum is $3.759\times10^{-42}~\rm{cm}^2$,
which is about 60 times the neutrino-electron elastic scattering process.
In addition, $\rm{{}^{37}Cl}$ also contributes about seven percent to the CC event rate.
The contained $\rm{{}^{35}Cl}$ and $\rm{{}^{6}Li}$ also make the delay-coincidence detection for electron antineutrinos possible.
The detector with LiCl water solution is basically a MeV-scale $\nu_e$ and $\bar\nu_e$ spectrometer.

We investigated the physical properties of LiCl and its water solution.
LiCl can be purified by recrystallization. The water solution has an attenuation length of $11\pm1$~m.
A very high molarity of $\rm{{}^{7}Li}$ of 11~mol/L can be achieved for operations at room temperature.
The event rate of $\nu_e$ on $\rm{{}^{7}Li}$ in a LiCl water solution is comparable to the elastic scattering.
A compact detector proposal is made, and its energy and angular resolutions are estimated.
The CC signals can be well separated from the elastic signals by a solar angle cut.

The kinetic energy after the interaction on $\rm{{}^{7}Li}$ directly reflects the neutrino energy.
It demonstrates clear advantages in studying the solar neutrino upturn effect, the light sterile neutrinos, and the Earth matter effect. The sensitivities in studying the solar neutrino upturn and light sterile neutrinos are reported for a detector with a 320-ton fiducial mass.

\section{Acknowledgement}
This work is supported in part by
the National Natural Science Foundation of China (Nos.~12141503 and 11620101004),
the Ministry of Science and Technology of China (No.~2018YFA0404102),
the Key Laboratory of Particle \& Radiation Imaging (Tsinghua University),
and the CAS Center for Excellence in Particle Physics (CCEPP).

\bibliographystyle{apsrev4-1}
\bibliography{Lithium}

\begin{thebibliography}{64}
\bibitem [{YeL()}]{YeLiang}%
  \BibitemOpen
  \href@noop {} {}\bibinfo {note} {Ye Liang and others, to be
  published}\BibitemShut {NoStop}%
\bibitem [{Luo()}]{Luowt}%
  \BibitemOpen
  \href@noop {} {}\bibinfo {note} {Wentai Luo and others, to be
  published}\BibitemShut {NoStop}%
\end{thebibliography}

\end{document}